\documentclass[varenna]{cimento}

\usepackage{graphicx}
\usepackage{amsmath}
\usepackage{amssymb}
\usepackage{aas_macros}

\title{Charged particles in magnetic fields and cosmic ray transport}

\author{P.~Blasi} %\orcid{0000-0003-2480-599X}%
\institute{Gran Sasso Science Institute, via F. Crispi 7–67100, L’Aquila, Italy \\ INFN/Laboratori Nazionali del Gran Sasso, via G. Acitelli 22, Assergi (AQ), Italy}

\begin{document}

\maketitle

\begin{abstract}
The theory of transport of charged particles in cosmic magnetic fields is at the very center of the investigation of non-thermal phenomena in the universe, ranging from our local neighborhood to supernovae, clusters of galaxies or distant active galaxies. It is crucial to understand how particles get energized to non-thermal energies as well as to describe their motion from the sources to an observer or to another location in the universe. Here I summarize some essential, basic aspects of the theory and discuss some topics in the theoretical framework that are currently being developed. I will also discuss some simple applications of the theory of transport to particle acceleration and propagation in the Galaxy. 
\end{abstract}

\section{Introduction}
\label{eq:intro}

In 1962 Bruno Rossi wrote his famous book on Cosmic Rays, in the occasion of the 50th anniversary of the discovery of cosmic rays (CRs). The book was published in 1964 \cite{rossi1964cosmic} and in the preface to the book, the author writes that most likely the historians of science will consider the field of cosmic rays closed with that anniversary. The transition of the field of cosmic rays into the birth crib of particle physics and its decline as a stand-alone research field, had been completed and particle physics was moving to terrestrial laboratories. The beginning of that transition can probably be traced back to the School held in Varenna in 1954, where Enrico Fermi lectured on the physics of pions. In hindsight, it may be interesting to notice that most of the topics discussed in the 2022 edition of the Varenna School {\it 'Foundations of Cosmic Ray Astrophysics'}, the first on topics related to CRs since 1954, could not even have been imagined in 1962: the basics of particle transport were laid down in the mid '60s to mid '70s and the discovery of diffusive shock acceleration dates back to the late 70's \cite{Krymskii1977,Bell1978,BlandfordOstriker1978} (see also reviews in Refs. \cite{Drury1983,BlandfordEichler1987}). By making use of many other disciplines (astrophysics, plasma physics, particle physics) and by contributing to them, the investigation of the origin of cosmic ray provided the seeds for what we now call {\it Astroparticle Physics}.

The lesson to learn is that, given this strong interplay with other fields, and given the new developments arising from ever better observations and measurements, the search for the origin of cosmic rays remains very dynamic. To this day, the study of particle acceleration, particle transport and radiation processes in different astrophysical setups remains one of the  most prolific in the field of Astrophysics.

Whether it is to describe the transport of cosmic rays in the Milky Way or in a distant galaxy, particle acceleration at a supernova shock or in the galactic wind of a starburst galaxy, the field rests quite solidly upon the pillar of particle transport in magnetic fields, which will be the topic of these lecture notes. 

One could dare and say that perhaps the whole field of high energy astrophysics reduces to three big questions: 1) how does Nature accelerate particles, namely how does it select some of the ambient plasma particles and energizes them to become non-thermal? 2) How do these non-thermal particles propagate in the complex environment inside the source, around the source and from the source to an observer? 3) How do these particles lose energy and radiate? 

Of these three big questions, the first two are certainly rooted in the description of charged particle transport in a non trivial magnetic field, independent of the specific environment that is being considered. 

The topic is vast and branches out into many subtopics, hence I had to make an effort here to select the points that to me seemed to be most important or most fundamental. By these terms I mean that the content of these Lecture Notes can be used to start attacking more detailed or more complex treatments of the same problem, and there is no lack of them. 

It is important to realize that the investigation of the physics of particle transport is all but concluded: this is in fact one of the fields in which we have experienced a prominent development, both in terms of phenomenology, in order to accommodate the high precision measurements that are becoming available, and in terms of fundamental principles, to reflect our improved understanding of turbulence and the ability of non-thermal particles to produce turbulence through the excitation of plasma instabilities. 

Yet at the same time, there are problems that are not as well understood as one would wish: for instance, we still do not know what is responsible for Galactic CR scattering at high energies, where self-generation fails and anisotropic development of MHD Alfvenic turbulence or damping of other modes may make MHD turbulence inadequate to provide a description of CR scattering. Another example of a field that is still lacking closure is the description of perpendicular diffusion, that still relies on phenomenological treatments  that often do not compare well with the results of numerical simulations of particle transport. These are just two instances, while more discussion on the answered questions in the field will be presented at the end of this manuscript. 

In these lecture notes, in addition to providing a description of the basic aspects of the theory of transport of charged particles in magnetic fields, I will discuss some noteworthy applications of the theory that will arise and emphasize some recent developments. 

The Notes are organized as follows: in Sec. \ref{sec:how} I will discuss some of the reasons why we think that the transport of non-thermal particles in magnetic fields is non-trivial; in Sec. \ref{sec:B0} I will briefly summarize the basic motion of a charged particle in an ordered magnetic field, while this description will be generalized to a more complex field in Sec. \ref{sec:deltaB}, where I first introduce the possibility that the motion may become diffusive in pitch angle and in physical space. In Sec. \ref{sec:vlasov} I will describe the motion of an ensemble of collisionless charged particles using the Vlasov equation and show that it leads to pitch angle diffusion. In Sec. \ref{sec:spatial} I obtain the equation describing spatial diffusion of charged particles, that plays such a central role in cosmic ray physics, as well as in describing a vast class of sources from clusters of galaxies to AGN and starburst galaxies, both in terms of simple transport and particle energization. In Sec. \ref{sec:applica} I apply the transport equation previously derived to two situations: {\it a)} the transport of Galactic CRs and {\it b)} diffusive shock acceleration at a newtonian shock. In Sec. \ref{sec:perp} I briefly outline the problem of diffusion in the direction perpendicular to the ordered magnetic field, mentioning some of the approaches that have been presented in the literature to this complex problem. In Sec. \ref{sec:nonlin} I sketch the essential aspects of the non-linear approaches to CR transport, in three contexts: {\it a)} diffusive shock acceleration, {\it b)} Galactic CR transport, {\it c)} transport of CRs around their sources, and {\it d)} transport of CR escaping the Galaxy and other galaxies as sources of CRs. I summarize the main points of these Lecture Notes in Sec. \ref{sec:concl}.

\section{How do we know that charged particles must perform a non trivial motion?}
\label{sec:how}

Most of the universe is in the form of a plasma, with very large electric conductivity. This condition implies that large scale electric fields are, in general, hard to support, with a few noteworthy exceptions (e.g. pulsar magnetospheres, solar flares and a few others), where locally the conditions of ideal MHD are broken and electric fields can be sustained. In such regions one can expect new phenomena to take place, such as reconnection. However, in most of the universe the absence of electric fields on large scales remains a valid approximation. It is worth noticing that magnetic fields do not make work on a charged particle, which is equivalent to say that they cannot change the energy of such particles. Hence, in the absence of electric fields, particle acceleration is not possible. 
However, the previous statement about the absence of electric fields in a plasma should be expressed more accurately: in fact the only electric fields that cannot be short circuited are the ones associated with the motion of a magnetized plasma. Such induced electric fields are, by definition, of strength $\epsilon\approx \frac{V}{c}B$, where $V$ is the velocity of plasma motion and $B$ is the strength of the magnetic field.  In most cases of interest here $V\ll c$, hence the resulting electric fields are small (obvious exceptions to this rule of thumb are relativistic plasmas, where the plasma motion and the speed of light become comparable). 
If the magnetic field and the plasma motion were coherent over a time scale T, during such time the equation of motion of the particle would be $\frac{dp}{dt}=q\frac{V}{c}B$, and upon integration, if the particle is assumed to be relativistic, this would lead to a maximum energy
\begin{equation}
E_{max}=q R B,
\label{eq:Hillas}
\end{equation}
where I introduced the size of the system $R=VT$. Eq. \ref{eq:Hillas} illustrates what is known as the Hillas criterion \cite{Hillas1984}. It can be easily rephrased in terms of the Larmor radius of the particles with maximum energy, $r_L(E_{max})=E_{max}/qB$ and reads $r_L(E_{max})=R$: the maximum energy is reached when the Larmor radius equals the size of the system. 

This rather intuitive criterion provides a very optimistic estimate of the maximum energy, which is unfortunately rarely satisfied in Nature, mainly because the induced electric field cannot be kept coherent over the age (or the spatial size) of the acceleration region. The Hillas criterion should be considered as an absolute upper bound to the maximum energy of a generic accelerator, but one should be very aware of the limitations of such a criterion. For instance, when applied to a supernova remnant shock, with a size of $R=10$ pc and a magnetic field $B=3\mu$G, the Hillas criterion would suggest $E_{max}\approx 3$ PeV, while dedicated calculations show that a typical SNR can hardly accelerate particles up to hundreds of TeV (see \cite{Blasi2013,Amato2014} for  reviews), and even in that case it requires the excitation of CR induced instabilities \cite{Bell1978,Lagage1983-1,Lagage1983-2,Bell2004,Cristofari2020,Cristofari2021}. 

Notice that even the optimistic Hillas criterion is based upon the hidden assumption that the particle is bound to stay in the acceleration region for at least a time $T=R/V$, in excess of the ballistic crossing time $t_b=R/c$. In other words, it requires that the particle motion is not a trivial gyration around magnetic field lines. This could be achieved by either a peculiar topology of the magnetic field line (for instance a closed line) or by frequent scattering of the particles, which would break the ballistic propagation. These considerations are general enough to be applicable to any acceleration region in which there is only a (non-relativistic) moving magnetic field.

An even more clear proof that the particle motion in a magnetized plasma is all but trivial is provided by propagation of cosmic ray nuclei in the Galaxy. It is commonly accepted that the particles that become CRs belong to either the interstellar medium (for instance for particles accelerated at the forward shock of a supernova remnant) or to stellar matter (as in the case of particle acceleration at stellar winds' termination shocks). Hence elements such as carbon and oxygen, abundant in the ISM and in stars, are considered as primary elements (for the most part). On the other hand, elements such as boron, lithium and beryllium have a very low abundance in that they could not be synthesized during big bang nucleosynthesis (the universe became colder and more rarefied faster than the reactions leading to the production of these elements from H and He could take place). Moreover they are destroyed in stars faster than they are produced. Hence the ISM is depleted of these elements. When the abundance of these elements is measured in CRs \cite{Munoz1987}, they are only slightly less abundant (within factors of order unity) than primary elements, such as C and O (see \cite{AMS-secondary-2018} for recent AMS-02 measurements). The accepted explanation of this finding is that B, Be, Li (as well as the so-called sub-Iron elements) are formed in spallation processes of primary elements. Since the cross sections for these processes are relatively well known, $\sigma_{sp}\approx 45 \rm A^{0.7}~\rm mb$, where $A$ is the mass number of the parent nucleus, one can estimate that the observed ratio of secondary to primary fluxes (for instance B/C or B/O) require that the primary nuclei remain confined within the Galactic volume for times not too far from the spallation time scale:
$$
\tau_{sp}\approx [n_d (h/H) \sigma_{sp} c]^{-1} \approx 100  H_4 A_{12}^{-0.7}~\rm Myr,
$$
where I assumed that the gas is distributed in a thin disc of half thickness $h\sim 150$ pc and the halo extends over a size $H=4 H_4$ kpc. The mass number is normalized here to that of C nuclei ($A=12$). 

A similar conclusion can be reached by studying some unstable isotopes: for instance $^{10}Be$, which has a half-life of 1.4 Myr, decays mainly to $^{10}$B during a time scale that is relativistically boosted by the Lorentz factor of the nucleus. The abundance of this isotope as compared with the stable $^9Be$ isotope is compatible with a propagation time of order several tens of million years. Since the decay of $^{10}Be$ mainly results in the production of $^{10}B$, also the ratio B/Be is sensitive to the propagation time (or to the size $H$ of the halo) and returns a similar estimate \cite{Evoli2020}. The time scales inferred from these exercises should be compared with the typical ballistic time scale over a distance $H$, $\tau_b\sim H/c\sim 10^4H_4$ years. The difference, about four order of magnitude, serves as the best indicator that the motion of charged particles in the Galaxy is severely modified with respect to ballistic motion. 

The theory of transport of charged particles in a non-trivial magnetic field addresses this problem and returns an elegant explanation of the reason why magnetic fields can confine particles in a given volume, be such volume the accelerator or the whole Galaxy. Below we discuss this theory from first principles.

\section{Motion of a charged particle in an ordered magnetic field}

\label{sec:B0}

I start the description of the motion of a charged particle in a magnetized plasma with the simplest scenario: the magnetic field is homogeneous, with a strength $B_0$ and is oriented along a direction I will call $\hat z$. In terms of components: $\vec B = (0,0,B_0)$. The particle is assumed to have a charge $q$ and the three-velocity is $\vec v = (v_x,v_y,v_z)$. Since there is no electric field in the system, the Lorentz force is simply $q \frac{\vec v}{c}\times \vec B$ and the equation of motion is 
\begin{equation}
\frac{d\vec p}{dt} = q \frac{\vec v}{c}\times \vec B\,.
\label{eq:Lorentz}
\end{equation}
In the absence of electric fields the energy of the particles stays constant in time, hence we know {\it a priori} that the Lorentz factor $\gamma=(1-v^2/c^2)^{-1/2}$ cannot depend on time. In other words, the modulus of the velocity vector stays constant. Recalling that $\vec p = m \gamma \vec v$, one can write the different components of Eq.~\ref{eq:Lorentz} as:
\begin{eqnarray}
m\gamma \frac{dv_x}{dt} = q\frac{v_y}{c} B_0\\
m\gamma \frac{dv_y}{dt} = -q\frac{v_x}{c} B_0\\
\frac{dv_z}{dt} = 0.
\end{eqnarray}
The third equation shows that in the absence of electric fields, the motion of the particle in the $z$-direction is rectilinear and uniform (there is no force in that direction), hence the $z$-component of the velocity vector is preserved both in modulus and direction. Upon introducing the pitch angle $\theta$ between the velocity vector and the $\hat z$ axis, and the pitch angle cosine $\mu=\cos(\theta)$, I can write $v_z=v\mu$, and since $v_z$ and $v$ are constant in time, it follows that $\mu$ is also constant in time. The pitch angle $\mu$ is a constant of motion in the presence of an ordered uniform magnetic field. 

Differentiating one of the first two equations above with respect to time and replacing the result in the other equation results in 
\begin{equation}
\frac{d^2 v_{x,y}}{dt} = -\left( \frac{q B_0}{m c \gamma}\right)^2 v_{x,y}\equiv -\Omega^2 v_{x,y},
\label{eq:gyration}
\end{equation}
where, in the last step, I have introduced the gyrofrequency $\Omega=qB_0/mc\gamma$. Eq. \ref{eq:gyration} is the well known equation describing a periodic motion of the variables $v_x$ and $v_y$ in time, with a frequency that is exactly $\Omega$, that is related to the Larmor (or gyration) radius of the particle through $r_L=v/\Omega$.  

The solution of the equation for $v_x(t)$ and $v_y(t)$ can be written in the following general form:
\begin{eqnarray}
v_x(t)=A \cos (\Omega t) + B \sin (\Omega t)\\
v_y(t)=B \cos (\Omega t) - A \sin (\Omega t),
\end{eqnarray}
and requiring that at time $t=0$ the initial conditions are fulfilled, namely $v_x(0)=v_\perp \cos\phi$ and $v_y(0)=v_\perp \sin \phi$, one immediately infers that $A=v_\perp \cos\phi$ and $B=v_\perp \sin\phi$, where $\phi$ is an arbitrary phase with, in general, no physical significance, and $v_\perp=v(1-\mu^2)^{1/2}$ is the modulus of the  velocity vector in the $xy$ plane. 

The periodic orbit of the particle in the $xy$ plane can now be written as 
\begin{eqnarray}
v_x(t)=v_\perp \left[\cos\phi\cos (\Omega t) + \sin\phi\sin (\Omega t)\right] \equiv v_\perp \cos\left(\phi - \Omega t\right)\\
v_y(t)=v_\perp\left[ -\cos\phi\sin (\Omega t) + \sin\phi\cos (\Omega t)\right]\equiv v_\perp \sin\left(\phi - \Omega t\right).
\end{eqnarray}
These expressions, together with $\mu=\rm constant$ completely define what I will refer to as the unperturbed trajectory, namely the particle trajectory in the presence of only an ordered magnetic field. 

Clearly the gyration around $\vec B_0$ and the rectilinear motion along $\vec B_0$ do not solve the problem with time scales discussed in the previous section, in that the motion along the field is still occurring at an appreciable fraction of the ballistic velocity, unless $\mu$ is very close to zero. This however would apply only to a negligibly small fraction of the particles. 

In the following sections I will discuss what happens when, in addition to the regular field $\vec B_0$, there are also small perturbations. The perturbative theory that I will discuss will allow us to compute the effect of the small perturbations to the unperturbed trajectory. I will proceed in two steps: the first, more phenomenological, will allow us to build a physical perception of how the motion of the particle changes in the presence of small magnetic perturbations. The second, more formal, will allow us to derive a transport equation for particles in a turbulent magnetic field. 

\section{Motion of a charged particle in a magnetic field $\vec B=\vec B_0+\delta \vec B$}
\label{sec:deltaB}

\subsection{Pitch angle diffusion}
\label{sec:pitch}

In this section I consider the possibility that, in addition to the ordered magnetic field $\vec B_0=B_0 \hat z$, there are small perturbations that, for simplicity, I assume to be only in the plane perpendicular to the $\hat z$ axis: $\delta\vec B=(\delta B_x,\delta B_y,0)$. This is the case if the perturbations are in the form of Alfv\'en waves propagating along $\vec B_0$. Throughout these notes I will assume that the perturbations are small compared with the unperturbed quantities, so as to make it possible to develop a perturbative approach, hence, $\delta B\ll B_0$.

In this situation the equation of motion of the particle, $\frac{d\vec p}{dt}=q\frac{\vec v}{c}\times (\vec B_0 + \delta\vec B)$ reads:
\begin{eqnarray}
\label{eq:mot1}
m\gamma \frac{dv_x}{dt} =\frac{q}{c} \left[v_y B_0 - v_z \delta B_y\right],\\
\label{eq:mot2}
m\gamma \frac{dv_y}{dt} = \frac{q}{c}\left[-v_x B_0+v_z \delta B_x\right],\\
\label{eq:mot3}
m\gamma \frac{dv_z}{dt} = \frac{q}{c}\left[ v_x \delta B_y - v_y \delta B_x \right],
\end{eqnarray}
where I again assumed that there are no electric fields. In fact this assumption requires now some more discussion: as mentioned above, the perturbations of the magnetic field can be considered as Alfv\'en waves (or other kinds of MHD modes). In this case, the magnetic field of the wave may be associated with an electric field, due to the motion of the wave. For instance, for Alfv\'en waves the electric field has a strength $\delta \epsilon=\frac{v_A}{c}B_0$, where $v_A=B_0/\sqrt{4\pi \rho}$ is the Alfv\'en speed of the wave in a medium with density $\rho$.  Since in typical conditions $v_A/c\ll 1$ (this would not be the case in a relativistic plasma), the electric field is small. Eqs. \ref{eq:mot1}-\ref{eq:mot3} can then be interpreted either as the approximate equations of motion in the assumption of negligibly small electric fields or, more correctly, the equations of motion of the particle in the reference frame in which the wave is at rest (the wave frame). This second interpretation is preferable in that, as discussed below, it retains the essential physical aspects that lead to second order Fermi acceleration. 

In the wave frame, the absence of electric fields ensures that the energy of the particle is conserved, namely $\gamma$ is constant in time, as assumed in Eqs. \ref{eq:mot1}-\ref{eq:mot3}. The RHS of Eqs. \ref{eq:mot1}-\ref{eq:mot2} contains a term proportional to $B_0$ and one proportional to the perturbation. This implies that the trajectory in the xy plane is only weakly perturbed. On the other hand, the RHS of Eq. \ref{eq:mot3} contains only perturbed quantities and this tells us that since $v$ is constant, the pitch angle cosine $\mu$ changes in time. In Eq. \ref{eq:mot3} I can replace the unperturbed solution for $v_x(t)$ and $v_y(t)$ as I calculated above:

\begin{equation}
m\gamma v\frac{d \mu}{dt} = \frac{q}{c}v_\perp \left[ \cos(\phi-\Omega t)\delta B_y - \sin (\phi-\Omega t) \delta B_x\right].
\label{eq:mu}
\end{equation}

In the wave frame we can imagine the perturbation to have a purely spatial wave dependence. Moreover we can assume that the perturbation is circularly polarized, namely $\delta B_y = \pm i \delta B_x$. Hence:
\begin{eqnarray}
\delta B_y = \delta B \exp[i(k z+\psi)]=\delta B\left\{ \cos(kz+\psi)+i \sin(kz+\psi)\right\}\\
\delta B_x = \delta B\left\{ \mp i\cos(kz+\psi) \pm \sin(kz+\psi)\right\},
\end{eqnarray}
where $\psi$ is an arbitrary phase. Since the fields are real quantities, we can take the real part and write:
\begin{eqnarray}
\delta B_y = \delta B\cos(kz+\psi)\\
\delta B_x = \pm \delta B \sin(kz+\psi).
\label{eq:polar}
\end{eqnarray}
Replacing these expressions in Eq. \ref{eq:mu} we obtain:
\begin{eqnarray}
m\gamma v\frac{d \mu}{dt} = \frac{q}{c}v_\perp \delta B\left[ \cos(\phi-\Omega t)\cos(kz +\psi) \pm \sin (\phi-\Omega t) \sin(kz+\psi)\right]\to\\
\to m\gamma v\frac{d \mu}{dt} = \frac{q}{c}v_\perp \delta B \cos(\phi-\Omega t \pm kz\pm \psi).
\label{eq:mu1}
\end{eqnarray}

Using the unperturbed trajectory we can write $z\approx v\mu t$, so that 

\begin{equation}
    \frac{d \mu}{dt}=\frac{q B_0}{m c\gamma}\frac{\delta B}{B_0}(1-\mu^2)^{1/2}
    \cos\left[\phi\pm \psi \pm(kv\mu\mp\Omega) t\right].
\end{equation}
The periodic dependence on time leads to the immediate conclusion that once averaged over a long period of time, 
\begin{equation}
    \langle \Delta\mu\rangle = \int_0^T \frac{d \mu}{dt} dt = 0.
\end{equation}
Although the mean value of the displacement in the cosine of the pitch angle vanishes, this does not necessarily mean that nothing interesting is happening: in fact there are other phenomena in which this is the case. Think of a drop of ink in a shallow water container. Because of the random molecular collisions, the stain is bound to grow in size into roughly a circle (in ideal conditions): both the $x$ and $y$ coordinates with respect to the center of the circle keep averaging to zero at any given time, and yet the size of the stain gets larger, namely $x^2$ and $y^2$ both increase in time. Not only they increase but it can be shown that they increase proportional to time, a distinctive feature of a diffusive (brownian) motion. 

In the case of a charged particle moving in a regular magnetic field plus small perturbations, there are no collisions because, in most astrophysical plasma of interest for us, the particle-particle collision rate is exceedingly small. There are however the random interactions of the charged particles with the magnetic perturbations. The mean value of the square of the pitch angle change can be written as:
$$
\langle \Delta\mu\Delta\mu\rangle = \left(\frac{q B_0}{m c\gamma}\right)^2 \left(\frac{\delta B}{B_0}\right)^2(1-\mu^2)\int_0^{2\pi} \frac{d\phi}{2\pi}\times 
$$
\begin{equation}
\times \int_0^Tdt\int_0^T dt' 
\cos\left[\phi\pm \psi \pm(kv\mu\mp\Omega) t\right]
\cos\left[\phi\pm \psi \pm(kv\mu\mp\Omega) t'\right],
\end{equation}
where we also averaged over the phase $\phi$ in the particle initial velocity. Using the identity $\frac{1}{2\pi}\int_0^{2\pi} \cos(\phi-a)\cos(\phi-b)=\frac{1}{2}\cos(a-b)$, and recalling the definition of gyrofrequency $\Omega$,  we obtain:
\begin{equation}
    \langle \Delta\mu\Delta\mu\rangle = \frac{1}{2}(1-\mu^2)\left(\frac{\delta B}{B_0}\right)^2\int_0^T dt \int_0^T dt' \cos\left[ (k v\mu\mp\Omega)(t-t') \right].
\end{equation}

In the limit $T\to \infty$ the integral converges to the definition of a Dirac delta function, and in such a limit we can write:
\begin{equation}
    \langle \Delta\mu\Delta\mu\rangle = \pi (1-\mu^2)\left(\frac{\delta B}{B_0}\right)^2
    T \delta(k v\mu\mp\Omega).
\label{eq:resonance}
\end{equation}
Eq. \ref{eq:resonance} is extremely important in that it shows in rather graphic way two essential facts about the propagation of a charged particle in a magnetized plasma with an ordered field and a perturbation of wavenumber $\vec k\parallel \vec B_0$: 1) the mean square value of the change in pitch angle vanishes unless the wavenumber equals the resonant wavenumber $k_{res}=\pm\Omega/v\mu$ for which the argument of the Delta function is zero. 2) The value of $\langle \Delta\mu\Delta\mu\rangle$ is proportional to $T$ and, as we discussed above, this is the distinctive feature of a diffusive behaviour. 

As we discuss in a more formal way below, it is customary to define a diffusion coefficient in pitch angle as
\begin{equation}
    D_{\mu\mu}=\frac{1}{2}\langle\frac{ \Delta\mu\Delta\mu}{T}\rangle = \frac{\pi}{2} (1-\mu^2)\left(\frac{\delta B}{B_0}\right)^2 \Omega
    |k_{res}|\delta(k - k_{res}).
\label{eq:Dmumu}
\end{equation}

It is instructive to analyze the physical meaning of Eq. \ref{eq:Dmumu}: as discussed above, the motion of the particle in the presence of perturbations in the magnetic field is diffusive if the argument of the delta function vanishes, which implies that the particle undergoes some sort of 'collision' (the analogous of brownian motion of a molecule of ink in water or the similar phenomenon of heat transfer). In our case no actual collision between particles is taking place. Rather the particles are changing pitch angle by interacting with perturbations in the magnetic structure of the plasma. However this motion manifests itself only if the resonance can be achieved. For $\mu\sim 1$ the resonance condition reads $k\sim \Omega/v=1/r_L(p)$: if the $k^{-1}$ of the perturbations is much smaller than the Larmor radius $r_L(p)$ of the particle with momentum $p$ then the perturbations average to zero over one Larmor gyration and the motion of the particles is only weakly perturbed. If on the other hand $k^{-1}$ is much larger than the Larmor radius, the particle can surf the small perturbation while not appreciably changing $\mu$. It is only when $k^{-1}\sim r_L$ that the resonance condition is satisfied and the particle motion becomes diffusive. 

Note that the resonance condition in Eq. \ref{eq:Dmumu} depends on $\mu$ and on the polarization of the wave. Hence, for instance, for positive $k$ the particles with $\mu<0$ ($\mu>0$) can only resonate with waves with the the minus (plus) sign in the polarization (see Eq. \ref{eq:polar}). We will discuss this further when we present our more formal description of the diffusive motion (sec. \ref{sec:vlasov}). 

An important point to keep in mind is that for $\mu\to 0$ the resonance wavenumber tends to infinity, where any well-behaved spectrum of perturbations has negligible power, thereby raising the issue of how does a particle cross through $\mu=0$ while scattering to eventually reverse its direction of motion along the $\hat z$ axis, an essential aspect of spatial diffusion. This aspect as well will receive more attention below in these Lecture Notes. 

\subsection{From pitch angle to spatial diffusion}
\label{sec:mu_to_space}

In Nature it is often the case that perturbations have a wide spectrum in $k$, namely there are components of the perturbed field on (almost) all scales. In fact this is the case even whenever the medium is stirred on one scale (or a narrow range of scales), because of cascading/transfer of power to smaller spatial scales. In fact, this topic represents one of the most active lines of current investigation, in that it is crucial to characterize the level of spatial diffusion that cosmic rays experience in the Galaxy. In Eq. \ref{eq:Dmumu}, the quantity $k_{res} (\delta B/B_0)^2$ can be interpreted as the amount of power available at the resonant scale. More formally, if power is available over a broad range of wavenumbers, Eq. \ref{eq:Dmumu} reads:
\begin{equation}
    D_{\mu\mu} = \frac{\pi}{2} (1-\mu^2) \Omega \int dk \left(\frac{\delta B(k)}{B_0}\right)^2 
    |k_{res}|\delta(k - k_{res})=\frac{\pi}{2} \Omega (1-\mu^2){\cal F}(k_{res}),
\label{eq:DmumuSpec}
\end{equation}
where we introduced the dimensionless power spectrum ${\cal F}(k_{res})=|k_{res}|(\delta B(k_{res})/B_0)^2$ at the resonant scale. The assumption that the perturbations are small and, as a consequence, the trajectory of the particle is only weakly perturbed, implies that ${\cal F}(k_{res})\ll 1$ at all scales. For transport in the Galaxy, as discussed below, this is a reasonably good approximation, while it fails in the description of transport in an acceleration region, for instance at SNR shocks, where the condition ${\cal F}(k_{res})\gg 1$ is required at least on some scales, if to accelerate particles up to very high energies, $\sim 100-1000$ TeV. It is clear that this inference implies that the perturbative approach proposed here should be used with a lot of caution in such a situation. 

With this in mind, it is clear from Eq. \ref{eq:DmumuSpec} that the larger is the power available at the resonant scale, the larger is the diffusion coefficient in pitch angle, namely diffusion is more efficient (notice that the opposite trend will appear in spatial diffusion).

Let us now address the issue of what this entails in terms of spatial diffusion of non-thermal particles. Here I will provide a phenomenological, yet very useful, view of this issue, while a formal calculation of spatial diffusion will be presented in Sec. \ref{sec:vlasov}. 
Since $d\mu=d\cos(\theta)=(1-\mu^2)^{1/2}d\theta$, one can easily write the diffusion coefficient in angle (rather than in $\mu$) as
\begin{equation}
    D_{\theta\theta}=\frac{1}{2}\langle\frac{ \Delta\theta\Delta\theta}{T}\rangle = \frac{\pi}{2} \Omega {\cal F}(k_{res}),
\label{eq:Dtheta}
\end{equation}
which has dimensions of radians/time. One can then estimate the time necessary for a change of angle by order one radian as 
\begin{equation}
\tau_{rev}\sim \frac{1}{D_{\theta\theta}}=\frac{2}{\pi} \Omega^{-1} {\cal F}^{-1}(k_{res}),
\label{eq:tau90}
\end{equation}
and since $\Omega^{-1}$ is the time necessary for one gyration and ${\cal F}\ll 1$, it follows that changing the direction of motion in an appreciable way requires numerous gyrations of the particle. 

Moreover, $\lambda_D=v\tau_{rev}$ can be interpreted as the pathlength for diffusion, namely the distance traveled by a particle on average before a reversal of direction (approximated here as one radian). Hence the diffusion coefficient in space, that can be generically written in terms of the pathlength is
\begin{equation}
    D_{zz}(p)=\frac{1}{3}v\lambda_D \approx 
    \frac{2}{3\pi}v^2\frac{1}{\Omega {\cal F}} = \frac{2}{3\pi} \frac{r_L(p)}{{\cal F}}v ,
\end{equation}
where we assumed that ${\cal F}$ is calculated at $k=1/r_L$ (as we will see in Sec. \ref{sec:vlasov}, an integration over $\mu$ is in order for a proper account of the dependence of $D_{\mu\mu}$ on $\mu$). Within factors of order unity, this expression for the diffusion coefficient coincides with the one we will derive later in a more formal way. If we introduce the so-called Bohm diffusion coefficient as $D_B(p)=\frac{1}{3}r_L(p) v$, the spatial diffusion coefficient becomes:
\begin{equation}
    D_{zz}(p)=\frac{2}{\pi} \frac{D_B(p)}{{\cal F}},
    \label{eq:Dzz}
\end{equation}
and since ${\cal F}\ll 1$, $D_{zz}$ is bound to be larger than the Bohm diffusion coefficient. Formally, diffusion becomes Bohm-like only when ${\cal F}(k)=1$ at all scales. The physical meaning of the Bohm diffusion regime is that the particle suffers a scattering each Larmor gyration, a situation that is however outside the region of applicability of a perturbative approach as the one we are discussing here. 

\subsection{Some phenomenological implications of these findings}
\label{sec:pheno}

We have just shown that in the presence of small perturbations in the magnetic field, the particle motion from ballistic (rectilinear uniform motion around $\hat z$ and circular motion in the $xy$ plane) becomes diffusive in both pitch angle and space. It is instructive to show the phenomenological implications of this apparently small effect, for instance in terms of Galactic CR transport.  

In fact one of the most convincing pieces of evidence that the CR motion in a magnetized plasma is non trivial is the measurement of the flux of secondary elements (such as B, Be and Li) \cite{Munoz1987} and the relative abundance of unstable to stable isotopes such as $^9$Be/$^{10}$Be, or other ratios such as B/Be \cite{AMS-secondary-2018}. From these observations, one can infer a size of the halo $H>5$ kpc and a confinement time $\sim H^2/D$ at $\sim 1$ GeV of order $\sim 80$ Myr \cite{Evoli2020,Evoli2019}. Using the expression for the spatial diffusion coefficient in Eq. \ref{eq:Dzz} and using $B_0=1\mu G$ to evaluate the Larmor radius, one obtains ${\cal F}(1 GeV)\simeq 10^{-6}$. In other words it is sufficient to have a perturbation of one part on a thousand in $\delta B/B_0$ (${\cal F}$ measures the power at that scale, proportional to the square of the perturbation) to transform the ballistic motion of the particle (crossing time $H/c\sim 10^4$ yr) into diffusive motion (diffusion time of order $80$ Myr): the effect is perturbative on the motion but the implications are macroscopic. 

For the purpose of establishing a contact between the theoretical aspects we are introducing and the phenomenology of CR transport in the Galaxy, it may be useful to ask whether the requirement of small amplitude of the perturbations is valid at all scales (namely at all energies). To address this issue we will make a few assumptions here for the sole purpose of reaching a quantitative assessment. We assume that the power spectrum of the perturbations is in the form $P(k)=P_0\left(\frac{k}{k_0}\right)^{-\alpha}$, where $k_0=1/L$ should be interpreted here as the inverse of the scale $L$ where the turbulence is injected and from where it eventually cascades toward smaller spatial scales (larger values of $k$). The normalization constant $P_0$ is evaluated by requiring that the total rms power is $\delta B^2_{tot}$, namely:
\begin{equation}
    \int_{k_0}^\infty dk P_0 \left(\frac{k}{k_0}\right)^{-\alpha} = \frac{P_0 k_0}{\alpha-1}=\left(\frac{\delta B_{tot}}{B_0}\right)^2 \to P_0=\frac{\alpha-1}{k_0}\left(\frac{\delta B_{tot}}{B_0}\right)^2.
\end{equation}
At this point we can compute the dimensionless power spectrum: 
\begin{equation}
    {\cal F}(k)= k P(k) = (\alpha-1)
    \left(\frac{k}{k_0}\right)^{1-\alpha} \left(\frac{\delta B_{tot}}{B_0}\right)^2,
    \label{eq:Fk}
\end{equation}
and require that at 1 GeV ($k=k_{res}=1/r_L(1 GeV)$) the dimensionless power is $\sim 10^{-6}$ (see above), so that
\begin{equation}
    \left(\frac{\delta B_{tot}}{B_0}\right)^2=\frac{10^{-6}}{\alpha-1}\left(\frac{k_{res}}{k_0}\right)^{\alpha-1}.
\end{equation}
It is customary to consider two cases of turbulent development in the ISM, one corresponding to the so-called Kolmogorov phenomenology, for which $\alpha=5/3$, and the other corresponding to the Kraichnan phenomenology, with $\alpha=3/2$. In passing it is useful to comment on the fact that using Eqs. \ref{eq:Dzz} and \ref{eq:Fk}, these two cases would correspond respectively to scaling of the diffusion coefficient with energy $D_{zz}(E)\propto E^{1/3}$ and $D_{zz}(E)\propto E^{1/2}$ res.

For $k_0^{-1}=L\approx 10$ pc, these two cases would lead to $ \left(\frac{\delta B_{tot}}{B_0}\right)^2\simeq 3\times 10^{-2}$ and $ \left(\frac{\delta B_{tot}}{B_0}\right)^2\simeq 3.5\times 10^{-3}$ respectively. Being both these numbers appreciably less than unity ensures that the perturbative approach is appropriate at all energies, including energies for which $r_L(E)=L$ (for the parameters adopted here this happens at $\sim 1$ PeV). 

In other words, the transport of CRs in the Galaxy seems to be well described by the perturbative approach in which $\delta B/B_0$ is small, over an extended range of energies of interest for CR physics.

There are however several caveats to this statement, that may be useful to briefly discuss since they are seeds of research activity that is being carried out at this time. The first comment is about the assumption that the turbulent spectrum may be written as a simple power law $P(k)\propto k^{-\alpha}$: it has been shown that stirring the ISM at scale $k_0$ and waiting for the Alfvenic cascade to develop, results in a very anisotropic turbulence \cite{GS1994}, in which most of the power is in the modes with wavenumber perpendicular to the direction of $\vec B_0$. Since the scattering is due to resonances between $k_\parallel$ and the particle pitch angle, this phenomenon must result in a reduction of the scattering rate \cite{YL2002,YL2004} and hence an increase in the spatial diffusion coefficient. Other MHD modes, such as fast magneto-sonic modes have been found to have a more isotropic cascade and be more efficient at scattering particles \cite{YL2008}, although the possibility has been discussed that they may form weak shocks, so that their role as particle scattering centers remains subject of debate \cite{Kempski2022}.  

The second caveat to the line of reasoning illustrated above is that CRs may produce their own scatterers through the excitation of streaming instability \cite{KulsrudPearce1969}, in the form of Alfv\'en waves mainly propagating down the CR gradient on Galactic scales. This is a non-linear scenario of CR transport in which the equilibrium spectrum of CR and the diffusion coefficient determine each other, in that they are both outputs of the calculation (see \cite{Skilling1974,Skilling1975,Holmes1975} for early discussion of this phenomenon and \cite{Blasi2012,AloBla2013,AloBlaSer2015,Kempski2022} for recent analyses, applied to AMS-02 data). It has been proposed that at least at rigidities below a few hundred GV this mechanism may be the main one for the confinement of Galactic CRs. In fact, in this picture, the spectral hardening measured by PAMELA \cite{Pam-hard} and AMS-02 \cite{AMS-hard-p,AMS-hard-He} and recently confirmed by DAMPE \cite{DAMPE-hard} and CALET \cite{CALET-hard} may arise as a consequence of a transition in CR transport from scattering off self-generated turbulence to scattering off extrinsic turbulence \cite{Blasi2012,AloBla2013,AloBlaSer2015} (see also \cite{Kempski2022} for a discussion of possible problems of this picture).

\subsection{On neglecting the electric field carried by waves}
\label{sec:electric}

In general, the waves discussed above also carry an electric field, with an amplitude that is of order $\delta \epsilon=(v_A/c)B_0$. In the previous discussion we have neglected such a field, and as anticipated above this assumption can be justified in several different ways, some carrying more physical content than others: one could simply say that since $v_A/c\ll 1$ the effect is small, which is true but not particularly insightful. One could also say that the wave dependence is $\exp\left[ -i\omega t + i k z \right]$, and since $z=v\mu t$ and for an Alfv\'en wave $\omega=k v_A$, provided $v\mu \gg v_A$ the effect of the wave motion (associated with the induced electric field) is small and hence negligible. This is equivalent to say that for a particle moving with an appreciable fraction of the speed of light, the wave appears to be basically at rest, hence again its electric field is negligible. Finally one can formally carry out the calculation in the reference frame of the wave, namely where the electric field vanishes. In this last approach one can wonder {\it a posteriori} what the effect of the neglected electric field is. 

It remains true that the mean value of all perturbations is zero, therefore $\langle \delta\epsilon \rangle=0$: it is easy to understand that the mean value of the momentum change must also vanish, but this simply means that the particle can either gain or lose a small fraction of its momentum, in each interaction with a wave. The small fraction is such that $\Delta p/p\sim v_A/c$. Similar to what has been done for diffusion in pitch angle, one can introduce a diffusion coefficient in momentum space: 
\begin{equation}
    D_{pp}=\frac{1}{2}\langle \frac{\Delta p \Delta p }{\Delta t}\rangle \approx p^2 \left(\frac{v_A}{c}\right)^2 \frac{1}{T},
\end{equation}
where $T\sim \tau_{rev}$ is the time necessary for substantial change in angle, due to diffusion in the pitch angle $\mu$ (see Eq. \ref{eq:tau90}). Now, it is easy to estimate the time necessary to change the momentum of the particle by $\sim p$:
\begin{equation}
    \tau_{pp}=\frac{p^2}{D_{pp}}=\left(\frac{c}{v_A}\right)^2 T \gg T.
\end{equation}
It follows that the effect of neglecting the electric field of the waves is what is known as second order Fermi acceleration \cite{Fermi1949,Fermi1954}: it occurs while particles are scattered in pitch angle but the time necessary for appreciable change in the modulus of the particle momentum is $\sim (c/v_A)^2$ times longer than the time for deflection of the particle by about one radian. For this reason this mechanism, though very important from the conceptual point of view and from the historic perspective, it has typically little effect on CR phenomenology. The only exceptions to this conclusion are the very low energy part of the Galactic CR spectrum, where the confinement time in the Galaxy is so long that some moderate level of acceleration (so-called reacceleration) may occur, and the obvious case in which the Alfv\'en speed is comparable with the speed of light, as it may be the case in gamma ray bursts and other relativistic sources. 

\section{Vlasov equation and particle motion in a magnetic field}
\label{sec:vlasov}

In the section above, I have derived the diffusive motion of the particles in a phenomenological way. It is, I believe, a useful approach, in that it allows one to have an insight into the problem without engaging in lots of mathematical details. On the other hand, we are left with the problem of finding an equation that describes the diffusive motion of charged particles both in pitch angle and in space. Moreover, the derived expressions for the diffusion coefficient in $\mu$ and in the $z$ coordinate are appropriate only as orders of magnitude but cannot be used in a proper, quantitative calculation of these phenomena. 

Here I adopt a more formal approach, based on the well known Vlasov equation, describing the behaviour of an ensemble of charged particles in a magnetized environment. The equation is, in fact, much more general than that, since it may easily include the effects of forces other than the Lorentz force (such as gravity). The approach presented here also has several points in common with the related approach (see Lecture Notes by A. Marcowith) aimed at the investigation of the dispersion relation of perturbations that are allowed in a plasma, perhaps in the presence of non-thermal particles. This latter study is of the utmost importance in that it is the basis of CR induced plasma instabilities that may explain magnetic field amplification at shocks, as well as self-confinement of CRs on Galactic scales, and many other phenomena (see Sec. \ref{sec:nonlin} for a brief discussion).  
The Vlasov equation can, at first sight, be interpreted as a continuity equation in phase space, or a mathematical formulation of the Liouville theorem. Let us introduce the density of non-thermal particles in phase space as $f(\vec p,\vec x,t)$, so that the density of such particles at the location $\vec x$ at time $t$ is
\begin{equation}
    n(\vec x,t) = \int d^3\vec p~ f(\vec p,\vec x,t).
\end{equation}
In the presence of only the Lorentz force, and assuming again that the electric fields are negligible, the Vlasov equation reads:
\begin{equation}
    \frac{\partial f}{\partial t} + \vec v\cdot \vec\nabla f+\frac{q}{c}\left(\vec v\times \vec B\right)\cdot \frac{\partial f}{\partial \vec p} = 0,
    \label{eq:vlasov}
\end{equation}
where the momentum $\vec p$ of the particles is related to their velocity $\vec v$ by $\vec p = m\vec v \gamma$. The magnetic field that appears in Eq. \ref{eq:vlasov} is the total field in the environment, which may be the sum of an ordered field $\vec B_0$ and perturbations. The field is described by Maxwell equations with source terms in the form of charge density and current density, which take into account both the background plasma and the non-thermal particles. In this sense, Eq. \ref{eq:vlasov} is an extremely powerful tool in that in principle it also describes the motion of an ensemble of particles under the action of self-generated perturbations. Here I will not discuss explicitly the origin of the perturbations (unless necessary) and I will simply assume that the magnetic field is $\vec B = \vec B_0+\delta \vec B$. Below I loosely follow the approach first introduced in some pioneering articles \cite{Volk1973,LeeVolk1975,Volk1975}. 

\begin{figure}
\centering
\includegraphics[width=1\textwidth]{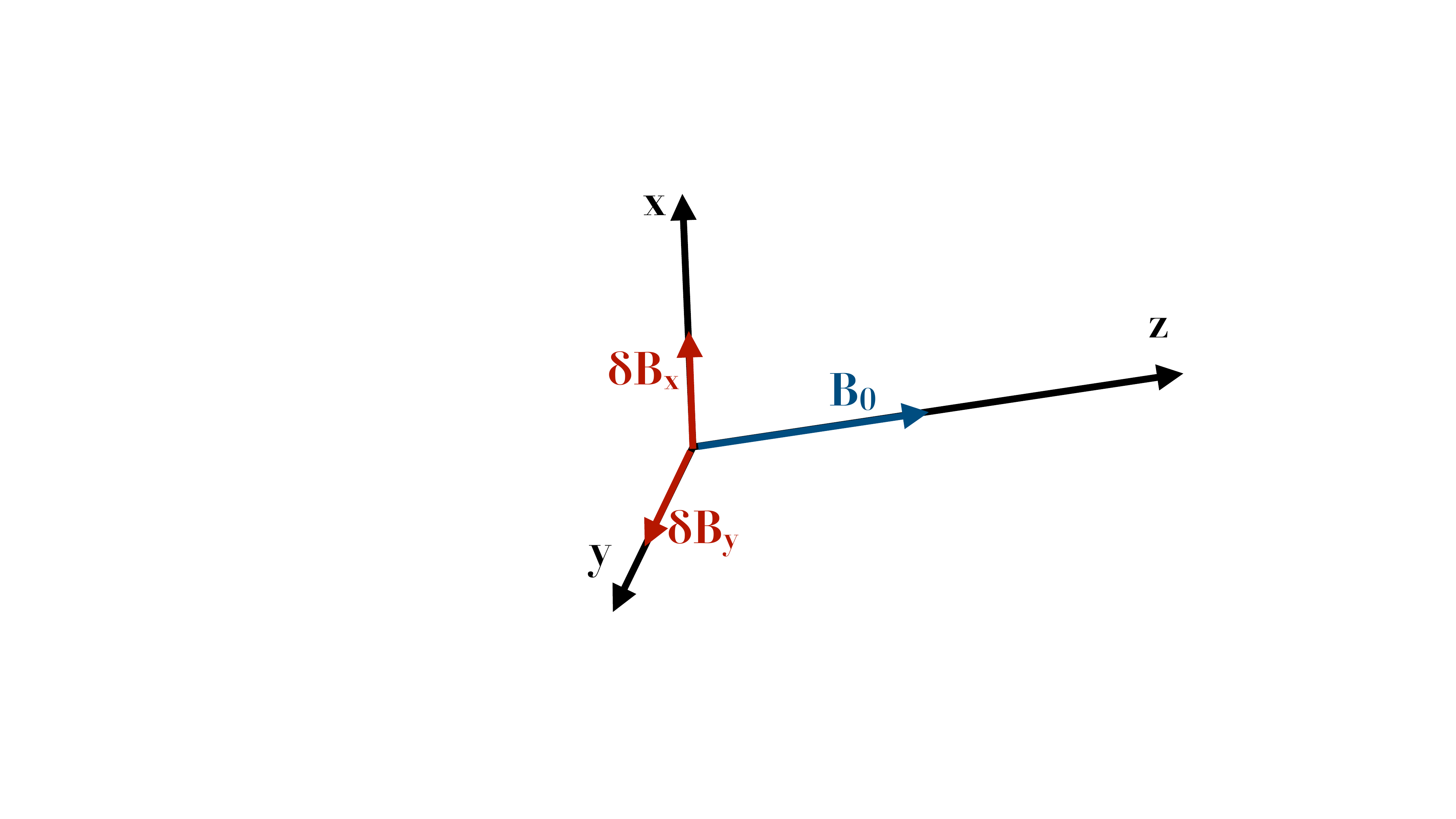}
\caption{Perturbed magnetic field in the $xy$ plane and $\vec B_0=B_0 \hat z$.}
\label{fig:geometry}
\end{figure}

I introduce the distribution function $f_0$ as the solution of Eq. \ref{eq:vlasov} when $\vec B=\vec B_0$. As in the previous section, with no loss of generality, I will assume that $\vec B_0=B_0\hat z$. If we focus, for simplicity, on perturbations that propagate along the direction of $\vec B_0$, then the perturbation is bound to be in the $xy$ plane (as a consequence of $\vec \nabla\cdot\vec B=0$). For an illustration of the geometry of the problem, refer to Fig. \ref{fig:geometry}.

The effect of perturbing the system is such that the distribution function will become $f=f_0+\delta f$, with $\langle \delta f \rangle=0$ (and as usual $\langle \delta \vec B \rangle=0$), to be interpreted as ensemble averages. Perturbing Eq. \ref{eq:vlasov} and retaining only terms that are at most first order in the perturbation (and recalling the definition of $f_0$) one obtains:
\begin{equation}
        \frac{\partial \delta f}{\partial t} + \vec v\cdot \vec\nabla \delta f+\frac{q}{c}\left(\vec v\times \vec B_0\right)\cdot \frac{\partial \delta f}{\partial \vec p} + \frac{q}{c}\left(\vec v\times \delta \vec B\right)\cdot \frac{\partial f_0}{\partial \vec p} = 0.
        \label{eq:vlasov_per}
\end{equation}
Given the geometry of the problem, the third term in Eq. \ref{eq:vlasov_per} becomes:
\begin{equation}
    \frac{q}{c}\left(\vec v\times \vec B_0\right)\cdot \frac{\partial \delta f}{\partial \vec p} = \frac{q}{c}B_0 \left( v_y \frac{\partial \delta f}{\partial p_x}-v_x \frac{\partial \delta f}{\partial p_y}\right).
    \label{eq:third}
\end{equation}

In a similar way the fourth term is: 
\begin{equation}
    \frac{q}{c}\left(\vec v\times \delta \vec B\right)\cdot \frac{\partial f_0}{\partial \vec p} = \frac{q}{c} \left[ -v_z \delta B_y \frac{\partial f_0}{\partial p_x}+v_z \delta B_x \frac{\partial f_0}{\partial p_y} + \left( v_x \delta B_y-v_y \delta B_x\right)\frac{\partial f_0}{\partial p_z}\right].
    \label{eq:fourth}
\end{equation}
Since the oredered field introduces a preferred direction, it is natural to adopt a cylindrical coordinate system so that
\begin{equation}
    v_x=v_\perp \cos \phi,~~~~~v_y=v_\perp \sin \phi,~~~~~v_z=v_\parallel,
\end{equation}
and similar for the momenta, so that
\begin{eqnarray}
dp_\perp = \cos\phi dp_x +\sin\phi dp_y\\
d\phi = -\frac{dp_x}{p_\perp}\sin\phi+\frac{dp_y}{p_\perp}\cos\phi.
\end{eqnarray}
Transforming the coordinate system in Eqs. \ref{eq:third} and \ref{eq:fourth} we get:
\begin{equation}
    \frac{q}{c}\left(\vec v\times \vec B_0\right)\cdot \frac{\partial \delta f}{\partial \vec p}=-\Omega \frac{\partial \delta f}{\partial \phi},
    \label{eq:term1}
\end{equation}
and
$$
    \frac{q}{c}\left(\vec v\times \delta \vec B\right)\cdot \frac{\partial f_0}{\partial \vec p}=
$$
\begin{equation}    
    =\frac{q}{c} \left[ -v_\parallel \delta B_y \cos\phi\frac{\partial f_0}{\partial p_\perp}+v_\parallel \delta B_x \sin\phi \frac{\partial f_0}{\partial p_\perp} + v_\perp\left( \cos\phi \delta B_y-\sin\phi \delta B_x\right)\frac{\partial f_0}{\partial p_\parallel}\right],
\end{equation}
where we explicitly assumed that the unperturbed distribution function $f_0$ is independent of the phase $\phi$ (namely $\partial f_0/\partial \phi=0$). Given the geometry of the problem this seems reasonable. It is however worth mentioning that this assumption implies that within this formalism we can only obtain diffusive motion in the parallel direction while perpendicular diffusion does not arise \cite{Volk1973,Volk1975}. 

At this point it is useful to introduce the two combinations of fields 
\begin{equation}
    \delta B_+=\delta B_x + i \delta B_y,~~~~~~~\delta B_-=\delta B_x - i \delta B_y,
\end{equation}
which can be inverted to
\begin{equation}
    \delta B_x=\frac{\delta B_++\delta B_-}{2}, ~~~~~~~ 
    \delta B_y=\frac{\delta B_+-\delta B_-}{2i}.
\end{equation}
In the formalism adopted earlier (Sec. \ref{sec:pitch}) in which $\delta B_y=\pm i \delta B_x$, the plus (minus) sign corresponds to a case where only the $-$ ($+$) polarization exists. In terms of these two polarization states the previous result reads:
\begin{equation}
     \frac{q}{c}\left(\vec v\times \delta \vec B\right)\cdot \frac{\partial f_0}{\partial \vec p} = i\frac{q}{2c} \left( v_\parallel \frac{\partial f_0}{\partial p_\perp}- v_\perp \frac{\partial f_0}{\partial p_\parallel} \right)\delta \tilde B,
     \label{eq:term2}
\end{equation}
where $\delta \tilde B=\delta B_+e^{-i\phi}-\delta B_-e^{i\phi}$.

At this point we can rewrite Eq. \ref{eq:vlasov_per} using the results in Eqs. \ref{eq:term1} and \ref{eq:term2}, to obtain:
\begin{equation}
    \frac{\partial \delta f}{\partial t} + \vec v\cdot \vec\nabla \delta f -\Omega \frac{\partial \delta f}{\partial \phi} = - \frac{i}{2c} \hat H f_0 \delta\tilde B,
\end{equation}
where I introduced the operator
\begin{equation}
    \hat H = v_\parallel \frac{\partial f_0}{\partial p_\perp}- v_\perp \frac{\partial f_0}{\partial p_\parallel}.
    \label{eq:vlasov_phi}
\end{equation}
In the following I will write again in an explicit form the expression for $\delta\tilde B$ since it contains the dependence on $\phi$, which we need to solve for. 

At this point I switch to Fourier space and express the perturbations through their Fourier transforms:
\begin{eqnarray}
    \delta B_i (z,t) = \int dk \int d\omega \delta B_i (k,\omega) \exp\left[ i k z - i \omega t\right]~~~~~~~i=x,y\\
    \delta f (z,t) = \int dk \int d\omega \delta f (k,\omega) \exp\left[ i k z - i \omega t\right].
\end{eqnarray}
In terms of the Fourier transforms, Eq. \ref{eq:vlasov_phi} is easily transformed to:
\begin{equation}
    -i \omega \delta f + i k v_\parallel \delta f - \Omega \frac{\partial \delta f}{\partial \phi} = -i \frac{q}{2c} \hat H f_0 \left( \delta B_+e^{-i\phi}-\delta B_-e^{i\phi}\right).
    \label{eq:deltaf}
\end{equation}
Since the dependence on the variable $\phi$ is only in the exponential terms, it is natural to look for solutions in the form $\delta f = C_1 e^{-i\phi}+ C_2 e^{i\phi}$. Replacement of this expression in Eq. \ref{eq:deltaf} returns 
\begin{equation}
    \delta f = \frac{q}{2c}\hat H f_0 \left[ \frac{\delta B_+}{\omega - k v_\parallel-\Omega}e^{-i\phi} - \frac{\delta B_-}{\omega - k v_\parallel+\Omega}e^{i\phi}\right].
    \label{eq:deltaf_phi}
\end{equation}
The physical meaning of this result is rather impressive: if we know the characteristics of the perturbed magnetic field, we also know how the distribution function is perturbed, as a function of $\phi$. This procedure (expressing $\delta f$ in terms of $\delta B$) is common to a very similar calculation that allows us to determine the dispersion relation for the perturbations that are allowed to propagate in the plasma, and which types of perturbations are either damped or grow exponential (instabilities), a topic covered in the Lecture Notes of A. Marcowith. 

Here I will use the relation between $\delta f$ and $\delta B$ in Eq. \ref{eq:deltaf_phi} to derive the equation describing pitch angle diffusion, one of the milestones of this field of investigation. 

To do so, I first consider again the Vlasov equation, Eq. \ref{eq:vlasov}, and calculate the ensemble average of both terms but retaining the second order terms, and recalling that $\langle \delta f\rangle = \langle \delta B\rangle = 0$:
\begin{equation}
    \frac{\partial f_0}{\partial t} + \vec v\cdot \vec\nabla f_0 = -\frac{q}{c} \langle\left(\vec v\times \vec \delta B\right)\cdot \frac{\partial \delta f}{\partial \vec p}\rangle. 
\end{equation}
This equation clearly shows how the evolution in space and time of $f_0$ is determined by the second order term on the RHS, where however $\delta f$ and $\delta B$ can be written in terms of the Fourier transforms in Eq. \ref{eq:deltaf_phi}, derived by considering only first order perturbations of the Vlasov equation. In a more mathematically explicit way this result can be written as:
$$
    \frac{\partial f_0}{\partial t} + \vec v\cdot \vec\nabla f_0 = 
$$
\begin{equation}
    -\frac{q}{c}
\int_0^{2\pi}\frac{d\phi}{2\pi} \int dk\int d\omega\int dk'\int d\omega' \langle\left(\vec v\times \vec \delta B (k',\omega')\right)\cdot \frac{\partial \delta f(k,\omega)}{\partial \vec p}\rangle,    
\label{eq:transport1}
\end{equation}
where, in addition to writing the perturbations in terms of their Fourier modes, I also averaged the result over the phase $\phi$ describing the initial phase of the motion of the particles. 

The RHS of Eq. \ref{eq:transport1} can be rewritten following a procedure that is very similar to the one adopted earlier for the first order perturbation terms: we need to write $\frac{\partial \delta f}{\partial \vec p}$ in cylindrical coordinates and recall that $\delta \vec B = (\delta B_x,\delta B_y, 0)$. After some simple algebra one then obtains:
\begin{eqnarray}
    \left(\vec v\times \vec \delta B\right)\cdot \frac{\partial \delta f}{\partial \vec p} = v_\parallel \frac{\partial \delta f}{\partial p_\perp}\left( -\delta B_y \cos\phi + \delta B_x\sin\phi\right)+ \nonumber\\
    \frac{v_\parallel}{p_\perp} \frac{\partial \delta f}{\partial \phi}\left( \delta B_y \sin\phi + \delta B_x\cos\phi\right)+ \nonumber\\
    v_\perp \frac{\partial \delta f}{\partial p_\parallel}\left( \delta B_y \cos\phi - \delta B_x\sin\phi\right).
\end{eqnarray}
If we introduce again the two states of polarization $\delta B_\pm$, this becomes:
\begin{eqnarray}
    \left(\vec v\times \vec \delta B\right)\cdot \frac{\partial \delta f}{\partial \vec p} = \frac{i}{2} \left( v_\parallel \frac{\partial \delta f}{\partial p_\perp} - v_\perp \frac{\partial \delta f}{\partial p_\parallel}\right)\left( \delta B_+ e^{-i\phi} - \delta B_- e^{i\phi} \right) + \nonumber\\
    +\frac{1}{2} \frac{v_\parallel}{p_\perp} \frac{\partial \delta f}{\partial \phi} \left( \delta B_+ e^{-i\phi} + \delta B_- e^{i\phi} \right).
    \label{eq:secondorder}
\end{eqnarray}
Notice that Eq. \ref{eq:secondorder} is in the form of an operator acting on the perturbation $\delta f$, and the latter has been previously calculated and has the form of Eq. \ref{eq:deltaf_phi}. Hence, replacing Eqs. \ref{eq:secondorder} and \ref{eq:deltaf_phi} in Eq. \ref{eq:transport1} and calculating the $\partial\delta f/\partial\phi$ using Eq. \ref{eq:deltaf_phi}, we can write:
\begin{eqnarray}
\frac{\partial f_0}{\partial t} + \vec v\cdot \vec\nabla f_0 = 
-i \left(\frac{q}{2c}\right)^2 \int_0^{2\pi}\frac{d\phi}{2\pi} \int dk\int d\omega\int dk'\int d\omega' \\
\langle \left\{ \left[ \left( \delta B_+(k',\omega') e^{-i\phi} - \delta B_-(k',\omega') e^{i\phi} \right) \hat H - \right.\right.\nonumber\\
\left.\left.\frac{v_\parallel}{p_\perp} \left( \delta B_+(k',\omega') e^{-i\phi} + \delta B_-(k',\omega') e^{i\phi} \right)\right] \right.\nonumber\\
\left.\left( \frac{\delta B_+(k,\omega)}{\omega - k v_\parallel-\Omega}e^{-i\phi} - \frac{\delta B_-(k,\omega)}{\omega - k v_\parallel+\Omega}e^{i\phi}\right) \hat H f_0\right\} \rangle.
\label{eq:transport2}
\end{eqnarray}
It is often assumed that the turbulence on which CR scatter is spatially homogeneous, which translates into the following condition for the correlator of fields:
\begin{equation}
    \langle \delta B_i (k,\omega) \delta B_j(k',\omega')\rangle = \frac{(2\pi)^4}{V T}  \langle \delta B_i (k,\omega) \delta B_j^*(k,\omega)\rangle \delta(k+k') \delta(\omega+\omega'),
\end{equation}
where $V$ and $T$ are a volume and a time scale sufficiently large that the turbulence can be considered homogeneous and stationary on those scales. 
This condition also tells us that since the fields are real, it must be realized that $\delta B_i(-k,-\omega)=\delta B_i^*(k,\omega)$, which in terms of $\delta B_\pm$ translates to
\begin{equation}
\delta B_+(-k,-\omega)=\delta B_-^*(k,\omega),~~~~~~~\delta B_-(-k,-\omega)=\delta B_+^*(k,\omega). 
\end{equation}
Using these definitions in Eq. \ref{eq:transport2} we easily handle it into:
\begin{eqnarray}
\frac{\partial f_0}{\partial t} + \vec v\cdot \vec\nabla f_0 = 
-i \left(\frac{q}{2c}\right)^2 \frac{(2\pi)^4}{V T}\int_0^{2\pi}\frac{d\phi}{2\pi} \int dk\int d\omega\\
\langle \left\{ \left[ \left( \delta B_-(k,\omega) e^{-i\phi} - \delta B_+(k,\omega) e^{i\phi} \right) \hat H - \right.\right.\nonumber\\
\left.\left.\frac{v_\parallel}{p_\perp} \left( \delta B_-^*(k,\omega) e^{-i\phi} + \delta B_+^*(k,\omega) e^{i\phi} \right)\right] \right.\nonumber\\
\left.\left( \frac{\delta B_+(k,\omega)}{\omega - k v_\parallel-\Omega}e^{-i\phi} - \frac{\delta B_-(k,\omega)}{\omega - k v_\parallel+\Omega}e^{i\phi}\right) \hat H f_0\right\} \rangle,
\label{eq:transport3}
\end{eqnarray}
and performing the simple integral over $\phi$ one promptly gets:

\begin{eqnarray}
\frac{\partial f_0}{\partial t} + \vec v\cdot \vec\nabla f_0 = 
i \left(\frac{q}{2c}\right)^2 \frac{(2\pi)^4}{V T} \int dk\int d\omega \nonumber\\
\left( \hat H + \frac{v_\parallel}{p_\perp}\right) \left[ \frac{\langle \delta B_-^* \delta B_-\rangle \hat H f_0}{\omega-kv_\parallel+\Omega}+\frac{\langle \delta B_+^* \delta B_+\rangle \hat H f_0}{\omega-kv_\parallel-\Omega}\right].
\label{eq:transport4}
\end{eqnarray}
One should notice that the LHS of Eq. \ref{eq:transport4} is a real quantity, which requires the RHS to be real too. Let us consider the term on the right, recalling that in general the frequency is a complex quantity $\omega=\omega_R+i\omega_I$. In the following I will assume that $\omega_R\gg \omega_I$ and $\omega_I>0$ (this last condition means that the waves are required to be weakly unstable). Therefore, I can write:
\begin{equation}
    \frac{i}{\omega_R+i\omega_I-k v_\parallel\pm\Omega}= \frac{i}{(\omega_R-k v_\parallel\pm\Omega)^2+\omega_I^2} + \frac{\omega_I}{(\omega_R-k v_\parallel\pm\Omega)^2+\omega_I^2}.
\end{equation}
We are now interested in evaluating the real part of this quantity in the limit $\omega_I\to 0$:
\begin{equation}
   \lim_{\omega_I\to 0} \frac{\omega_I}{(\omega_R-k v_\parallel\pm\Omega)^2+\omega_I^2} \equiv \pi \delta(\omega_R- k v_\parallel\pm\Omega).
\end{equation}
We can easily envision that the $\delta$-function that appeared in this result is the very reason why scattering shows a resonant nature. Its appearance is however the consequence of requiring that the waves are only weakly unstable, namely that the real part of their frequency is much larger than the imaginary part. There are situations in which this is not the case and the interaction between charged particles and waves is non-resonant. 

We notice now that, given a generic function $g$:
\begin{equation}
    \left( \hat H + \frac{v_\parallel}{p_\perp}\right)g=\frac{1}{p_\perp}\left( p_\perp\hat H + v_\parallel\right)g=
    \frac{1}{p_\perp}\left( p_\perp v_\parallel \frac{\partial}{\partial p_\perp}-p_\perp v_\perp\frac{\partial}{\partial p_\parallel} + v_\parallel\right)g.
    \label{eq:formula}
\end{equation}
It is useful to recall that
\begin{equation}
    v_\parallel \frac{\partial(p_\perp g)}{\partial p_\perp}=v_\parallel g + v_\parallel p_\perp \frac{\partial g}{\partial p_\perp},
\end{equation}
and replacing this in Eq. \ref{eq:formula}:
\begin{equation}
    \left( \hat H + \frac{v_\parallel}{p_\perp}\right)g = \frac{1}{p_\perp}\left(v_\parallel \frac{\partial}{\partial p_\perp}(p_\perp g) - p_\perp v_\perp \frac{\partial g}{\partial p_\parallel}\right)=\frac{1}{p_\perp}\hat H (p_\perp g).
\end{equation}
Using this result in Eq. \ref{eq:transport4}, we obtain the following more compact form for the Vlasov equation:
\begin{eqnarray}
\frac{\partial f_0}{\partial t} + v_\parallel \frac{\partial f_0}{\partial z} = 
\left(\frac{q}{2c}\right)^2 \frac{(2\pi)^4}{V T} \frac{\pi}{p_\perp}\int dk\int d\omega \nonumber\\
\hat H \left\{ p_\perp \left[ \langle \delta B_-^* \delta B_-\rangle \delta(\omega-k v_\parallel+\Omega) + \langle \delta B_+^* \delta B_+\rangle \delta(\omega-k v_\parallel-\Omega)\right] \right\}\hat H f_0.
\label{eq:transport5}
\end{eqnarray}
As mentioned several times through these notes, the waves we are referring to are Alfv\'en waves, although the treatment can be easily generalized to other MHD waves. In all these cases the frequency of the waves is low and it may be useful to take the {\it zero frequency limit} for the power spectrum:
\begin{equation}
    \lim_{V,T\to \infty} \frac{(2\pi)^4}{V T} \langle \delta B_\pm^* \delta B_\pm\rangle = \delta(\omega)P_\pm(k).
\end{equation}
With this assumption, Eq. \ref{eq:transport5} becomes:
\begin{eqnarray}
\frac{\partial f_0}{\partial t} + v_\parallel \frac{\partial f_0}{\partial z} = \nonumber\\
\left(\frac{q}{2c}\right)^2 \frac{\pi}{p_\perp}\int dk 
\hat H \left\{ p_\perp \left[ P_-(k) \delta(k v_\parallel-\Omega) + P_+(k) \delta(k v_\parallel+\Omega)\right] \right\}\hat H f_0.
\label{eq:transport6}
\end{eqnarray}
Notice that the assumption of vanishing frequency implies that there are no electric fields carried by the waves (it is as if we were sitting in the reference frame of the waves). Hence there cannot be changes in the modulus of the particle momentum, while the pitch angle can change. Taking this fact into account and switching to spherical coordinates, ($p_\parallel,p_\perp)\to (p,\mu)$, leads to:
\begin{equation}
    \frac{1}{p_\perp}\hat H \approx -\frac{v}{p^2} \frac{\partial}{\partial \mu},
\end{equation}
and replacing in Eq. \ref{eq:transport6}:
\begin{equation}
    \frac{\partial f_0}{\partial t} + v_\parallel \frac{\partial f_0}{\partial z} = \frac{\partial}{\partial \mu} \left[ D_{\mu\mu}\frac{\partial f_0}{\partial \mu} \right],
    \label{eq:transport_mu}
\end{equation}
where we introduced:
\begin{equation}
    D_{\mu\mu}=\frac{\pi}{4} \Omega^2 (1-\mu^2) \int dk \left[ \frac{P_-(k)}{B_0^2} \delta(k v_\parallel-\Omega) + \frac{P_+(k)}{B_0^2} \delta(k v_\parallel+\Omega)\right].
    \label{eq:Dmumu1}
\end{equation}
The reader might recall that the quantities $P(k)/B_0^2$ are the same dimensionless power spectra ${\cal F}(k)$ introduced in Sec. \ref{sec:mu_to_space}, so that Eq. \ref{eq:Dmumu1} can also be rewritten as: 
\begin{equation}
    D_{\mu\mu}=\frac{\pi}{4} \Omega^2 (1-\mu^2)\left[ {\cal F}_- \left(k=\frac{\Omega}{v\mu}\right)+{\cal F}_+ \left(k=-\frac{\Omega}{v\mu}\right)\right].
    \label{eq:Dmu_exact}
\end{equation}
These last few results are dense with physical meaning: first, we notice how the result in Eq. \ref{eq:Dmu_exact}, derived in a formal way contains essentially the same information about particle motion as derived earlier in these notes: the pitch angle changes as a response to resonances, and the order of magnitude of what was introduced as the diffusion coefficient in pitch angle is compatible with the approximate expression found in Eq. \ref{eq:DmumuSpec}. However it is noteworthy that we have now been able to derive the transport equation for the particles in pitch angle, Eq. \ref{eq:transport_mu}, namely to show that indeed the motion is diffusive, and that $D_{\mu\mu}$ plays the role of a diffusion coefficient. Eq. \ref{eq:Dmu_exact} also illustrates in a very clear way the role played by the wave polarization, since the power in the form of the two states has been written in a clear, separate way. 

The transport equation, Eq. \ref{eq:transport_mu}, has the form of a Boltzmann equation with a scattering term (RHS) that identifies the nature of the scattering as the resonant interactions between the gyromotion of the particles and the wavenumber of the waves. Notice that the resonance is sensitive to the sign of $\mu$: if the particles have positive electric charge ($\Omega>0$), if $k>0$, this forces the resonance to occur with the $-$ sign of the polarization if $\mu>0$ and with the $+$ sign of the polarization if $\mu<0$. 

\section{From diffusion in pitch angle to spatial diffusion}
\label{sec:spatial}

Eq. \ref{eq:transport_mu} describes the evolution in time $t$ and space $z$ of the distribution function $f_0$, given an initial condition. For instance one might consider applying such an equation to describe how a beam of particles gets broadened by the action of pitch angle scattering: we can envision the beam to get wider in $\mu$ until the momentum in the $\hat z$ direction has become zero and eventually the particles reverse direction of motion (if they manage to diffuse through $\mu=0$) and start moving in the opposite direction. This simple thought experiment illustrates how pitch angle diffusion must also result in spatial diffusion. 

In most applications that we will focus upon here we will assume that the particles are well scattered, namely that the resonant scattering is sufficiently efficient that the particle distribution is led to quasi-isotropy.  We will check {\it a posteriori} what this assumption entails. One can see that the first correction to isotropy of the distribution function is connected to dipole anisotropy and is strictly assumed here to be small.

Hence, it is clear that the approach to spatial diffusion that we will illustrate below is appropriate to the description of problems in which the anisotropy is small: this is the case for instance in the description of CR transport in the Galaxy or in other galaxies and starburst objects. It is also the case in the description of particle acceleration at newtonian shocks, for instance in supernova remnants and star clusters. On the other hand, it is not appropriate to describe particle acceleration at relativistic shocks, where the anisotropy both upstream and downstream of the shock is of order unity \cite{Vietri2003,BlasiVietri2005}. 

The derivation of the spatial transport equation discussed below follows closely the one presented by \cite{Shalchi_book}. 

Let us start from the transport equation in pitch angle, that we rewrite down here for simplicity:
\begin{equation}
    \frac{\partial f}{\partial t} + v_\parallel \frac{\partial f}{\partial z} = \frac{\partial}{\partial \mu} \left[ D_{\mu\mu}\frac{\partial f}{\partial \mu} \right],
    \label{eq:transport_mu1}
\end{equation}
where I adopted the symbol $f$ to denote the distribution function. Let us first introduce the moments of such distribution function. The first moment is the mean:
\begin{equation}
    f_0 (z,t)=\frac{1}{2}\int_{-1}^1 d\mu' f (z,t,\mu'),  
\end{equation}
where I omitted to indicate the dependence of these quantities on momentum since the particle energy cannot change in the absence of electric fields. Given the assumption that particles are nearly isotropic, $f_0$ is very close to the actual solution. I will refer to $f_0$ as the isotropic part of the distribution function. The deviations from isotropy are contained in the second moment, the current:
\begin{equation}
    J(z,t) = \frac{1}{2}\int_{-1}^1 d\mu' v\mu' f (z,t,\mu').
\end{equation}
One should notice that if $f=f_0$ then the current would vanish (there would be exactly the same number of particles flowing in the two directions along $\hat z$). 

If we now apply the operator $(1/2)\int_{-1}^1 d\mu'$ to Eq. \ref{eq:transport_mu1} we obtain:
\begin{equation}
    \frac{\partial f_0}{\partial t} + \frac{\partial J}{\partial z} = 0.
    \label{eq:cont}
\end{equation}
The RHS vanishes under the action of the operator because $D_{\mu\mu}\propto (1-\mu^2)$ (see Eq. \ref{eq:Dmu_exact}). Eq. \ref{eq:cont} is formally a continuity equation and tells us that the main part (the isotropic part) of the distribution function changes in time if there is a divergence of the current, which is intuitively easy to understand. 

We note that one can write $\mu=-\frac{1}{2}\frac{\partial}{\partial\mu}(1-\mu^2)$, hence:
\begin{equation}
    J=\frac{1}{2}\int_{-1}^1 d\mu' v\mu' f = - \frac{v}{4}\int_{-1}^1 d\mu' \frac{\partial}{\partial\mu}(1-\mu^2) f=
    \frac{v}{4}\int_{-1}^1 d\mu' (1-\mu^2) \frac{\partial f}{\partial \mu},
\end{equation}
where in the last step we used integration by parts. 

Let us now integrate Eq. \ref{eq:transport_mu1} between $\mu=-1$ and $\mu$:
\begin{equation}
\frac{\partial}{\partial t}\int_{-1}^\mu d\mu' f + \int_{-1}^\mu d\mu' v\mu' \frac{\partial f}{\partial z}=D_{\mu\mu}\frac{\partial f}{\partial \mu},
\end{equation}
and assuming $f\approx f_0$ in the firts two integrals:
\begin{equation}
\frac{\partial f_0}{\partial t} (\mu+1) + \frac{1}{2}(\mu^2-1) v \frac{\partial f_0}{\partial z}=D_{\mu\mu}\frac{\partial f}{\partial \mu}.
\end{equation}
Deriving $\partial f/\partial\mu$ from this equation and replacing it into the expression for the current leads to:
\begin{equation}
    J=\frac{v}{4}\int_{-1}^1 d\mu' (1-\mu^2) \left[ \frac{\partial f_0}{\partial t} \frac{(\mu+1)}{D_{\mu\mu}} + \frac{1}{2}\frac{(\mu^2-1)}{D_{\mu\mu}} v \frac{\partial f_0}{\partial z}\right] \equiv \kappa_t \frac{\partial f_0}{\partial t} - D_{zz} \frac{\partial f_0}{\partial z},
    \label{eq:temp}
\end{equation}
where I have introduced the two quantities:
\begin{eqnarray}
    \kappa_t = \frac{v}{4}\int_{-1}^1 d\mu' \frac{(1-\mu^2)(1+\mu)} {D_{\mu\mu}}\\
    D_{zz}=\frac{v^2}{8}\int_{-1}^1 d\mu' \frac{(1-\mu^2)^2} {D_{\mu\mu}}\label{eq:Dzz_exact}
\end{eqnarray}
In the first term on the RHS of Eq. \ref{eq:temp} I make use of the continuity equation, Eq. \ref{eq:cont}, so that 
\begin{equation}
    J=-\kappa_t \frac{\partial J}{\partial z} - D_{zz} \frac{\partial f_0}{\partial z}.
    \label{eq:J}
\end{equation}

If one consider to the lowest order in the anisotropy that $f\approx f_0+f_1 \mu$, with $f_1\ll f_0$, then 
\begin{equation}
    J\approx \frac{1}{2}\int_{-1}^1 d\mu' v\mu' (f_0+f_1 \mu') = \frac{1}{3}v f_1. 
\end{equation}
It follows that the first term in Eq. \ref{eq:J} is of order $(v^2/D_{\mu\mu})(f_1/z)$, while the second term is of order $(v^2/D_{\mu\mu})(f_0/z)$. Hence the second term is dominant and one can write:
\begin{equation}
   J\approx -D_{zz} \frac{\partial f_0}{\partial z},
   \label{eq:J-Der}
\end{equation}
and differentiating again with respect to $z$ and using the continuity equation:
\begin{equation}
    \frac{\partial f_0}{\partial t} = 
    \frac{\partial}{\partial z} \left[ D_{zz}\frac{\partial f_0}{\partial z}\right],
    \label{eq:transport_space}
\end{equation}
that is the sought after transport equation in space for the isotropic part of the distribution function. The quantity $D_{zz}$ in Eq. \ref{eq:Dzz_exact} plays the role of spatial diffusion coefficient, similar to the one derived in Sec. \ref{sec:pitch} in a more phenomenological way. 

Although Eq. \ref{eq:transport_space} describes the evolution of the isotropic part of the distribution function, as discussed above, the current carries information about the anisotropy, and Eq. \ref{eq:J-Der} clearly connects the anisotropy with the spatial derivative of the isotropic part of the distribution function. 

\subsection{The case of scattering in a moving plasma}
\label{sec:moving}

In many cases in astrophysical systems, transport occurs in a plasma that moves with respect to the laboratory (observer) frame. If the particles are "well scattered" in the comoving frame, their motion simply results in advection with the plasma, as it is intuitively clear. In other words, the net effect of a moving magnetized plasma is that an induced electric field appears in the observed frame, that causes a $\vec E \times \vec B$ drift: this is what we describe as advection with the plasma. This can be understood in terms of change of reference frame: if the velocity $u$ of the plasma is homogeneous, namely independent of position, then it is always possible to find a reference frame (comoving with the plasma) in which $u=0$ and the transport equation becomes again Eq. \ref{eq:transport_space}. The situation in which $u$ is not the same everywhere is clearly different in that there is no individual frame in which the velocity of the plasma vanishes. This is the interesting case that we want to consider next, in that it opens the way to applying the transport equation to problems such as diffusive shock acceleration. 

As discussed above, the transport equation in pitch angle can be considered as a generalization of the Boltzmann equation with a scattering term that is provided by the resonant interactions with waves. The LHS of Eq. \ref{eq:transport_mu1} can then be considered as the total derivative with respect to time. When the plasma in which diffusion takes place is itself in motion with respect to another frame, there are two ingredients that change: one is that the particle velocity is now $u+v\mu$ (where $u$ is assumed to be aligned with the $\hat z$ axis for simplicity and we assumed $u\ll c$, and the pitch angle $\mu$ is still measured in the comoving frame). The second thing that happens is that the energy $E$ and the momentum component in the $\hat z$ direction, $p_z=p\mu$ of a particle in the comoving frame, transformed into the laboratory frame become:
\begin{eqnarray}
    E'=\gamma E - \beta \gamma p\mu,\nonumber\\
    p'_z = -\beta\gamma\frac{E}{c}+\gamma p \mu,
    \label{eq:LorentzInv}
\end{eqnarray}
where $\beta=u/c\ll 1$ and $\gamma=(1-\beta^2)^{-1/2}$. The derivative of $f$ (which is a Lorentz invariant) with respect to $z$ becomes then
\begin{equation}
    \frac{\partial f}{\partial z} + \frac{\partial f}{\partial p'_z}\frac{\partial p'_z}{\partial z} = - \frac{\partial f}{\partial p'_z} \frac{d}{dz}\left( \frac{u}{c}\right) \frac{E}{c},
\end{equation}
where, on the RHS, we made use of the second of Eqs. \ref{eq:LorentzInv}.
Putting things together, the second term in Eq. \ref{eq:transport_mu1} becomes:
\begin{equation}
    (u+v\mu)\frac{\partial f}{\partial z} - \frac{E}{c^2} (u+v\mu) \frac{du}{dz} \frac{\partial f}{\partial p'_z} \approx (u+v\mu)\frac{\partial f}{\partial z} - \frac{E}{c^2} v\mu \frac{du}{dz} \frac{\partial f}{\partial p'_z},
\end{equation}
where in the last step we neglected terms of order $(u/c)^2$. It is useful to switch from cylindrical to spherical coordinates $(p'_\parallel,p'_\perp)\to (p',\mu')$, so that 
\begin{equation}
    \frac{\partial f}{\partial p'_z} = \frac{\partial f}{\partial p'_\parallel} = 
     \frac{\partial f}{\partial p'} \mu' +  \frac{\partial f}{\partial \mu'} \frac{1-\mu'^2}{p'}.
\end{equation}
Here $\mu'$ and $\mu$ are related through the Lorentz transformation. The generalization of the transport equation to the case of a plasma in motion is then easily written as:
\begin{equation}
    \frac{\partial f}{\partial t} + (v\mu+u) \frac{\partial f}{\partial z} - \frac{E}{c^2} v\mu \frac{du}{dz} \left[ \frac{\partial f}{\partial p'} \mu' +  \frac{\partial f}{\partial \mu'} \frac{1-\mu'^2}{p'} \right] = \frac{\partial}{\partial \mu} \left[ D_{\mu\mu} \frac{\partial f}{\partial \mu} \right].
    \label{eq:Boltzmann_motion}
\end{equation}
At this point we can again apply the operator $(1/2)\int_{-1}^1 d\mu$ to Eq. \ref{eq:Boltzmann_motion}, which leads to:
\begin{equation}
    \frac{\partial f_0}{\partial t} + u\frac{\partial f_0}{\partial z}+ \frac{\partial J}{\partial z} - \frac{1}{3} \frac{du}{dz} p \frac{\partial f_0}{\partial p'} = 0,
    \label{eq:Boltzmann_motion1}
\end{equation}
where we used the identity $Ev/c^2=p$ and as before:
\begin{equation}
    J=\frac{1}{2}\int_{-1}^1 d\mu f v \mu = \frac{v}{4} \int_{-1}^1 d\mu (1-\mu^2) \frac{\partial f}{\partial \mu}.
    \label{eq:Jfull}
\end{equation}
We can repeat the same steps as in the case of a plasma at rest and integrate Eq. \ref{eq:Boltzmann_motion} between $-1$ and $\mu$, to get:
\begin{equation}
\frac{\partial f_0}{\partial t} (1+\mu) + u \frac{\partial f_0}{\partial z} (\mu+1) +     \frac{1}{2}(\mu^2-1) v \frac{\partial f_0}{\partial z} - \frac{1}{6}\frac{du}{dz} p  \frac{\partial f_0}{\partial p} (\mu^3+1)=D_{\mu\mu}\frac{\partial f}{\partial \mu}.
\end{equation}
This expression can be used to derive $\partial f/\partial \mu$ to be used in Eq. \ref{eq:Jfull}. However, it can be easily seen that here, in analogy with what we have done in the simpler case of constant $u$, retaining the main terms only, one has:
\begin{equation}
    J\approx - D_{zz}\frac{\partial f}{\partial z},
\end{equation}
and taking another derivative with respect to $z$ and using Eq. \ref{eq:Boltzmann_motion1}, we get:
\begin{equation}
    \frac{\partial f_0}{\partial t} + u\frac{\partial f_0}{\partial z} - \frac{1}{3} \frac{du}{dz} p \frac{\partial f_0}{\partial p}
    =\frac{\partial}{\partial z}\left[D_{zz}\frac{\partial f_0}{\partial z}\right],
    \label{eq:transport_space_motion}
\end{equation}
which is the generalization of Eq. \ref{eq:transport_space} to the case in which the plasma is in motion with velocity $u$ in the $\hat z$ direction. One can notice that if $u$ is spatially constant, then the only effect is the obvious advection of CRs with the plasma (second term on the LHS). In fact $\frac{\partial}{\partial t}+u \frac{\partial}{\partial z}$ is the total time derivative operator. If however $du/dz\neq 0$ then in principle more interesting effects may arise, as we discuss below, as the energy of the particles can change. 

Although the transport equation was derived in one spatial dimension, it is possible to generalize its derivation to 3D:
\begin{equation}
    \frac{\partial f_0}{\partial t} + \vec u\cdot \vec\nabla f_0 - \frac{1}{3} (\vec\nabla\cdot\vec u) p \frac{\partial f_0}{\partial p}
    =\vec\nabla\left[D \vec\nabla f_0\right],
    \label{eq:transport_space_3D}
\end{equation}
where $f_0$ still denotes the isotropic part of the distribution function and we assumed for simplicity that the diffusion coefficient $D$ is a scalar, although for specific applications this might not be the case. In fact, as discussed in Sec. \ref{sec:perp}, the diffusion coefficient perpendicular to the magnetic field line, for weak turbulence, in which $\delta B/B_0<1$, is much smaller than the diffusion coefficient in the direction of $\vec B_0$. When this separation between the parallel and the perpendicular direction becomes important for the description of the phenomenological aspects of the problem at hand, Eq. \ref{eq:transport_space_3D} should be replaced with a more appropriate equation that retains the information about the direction of propagation (see for instance \cite{Shalchi_book}). 

\section{Two applications of the transport equation}
\label{sec:applica}

Eq. \ref{eq:transport_space_motion} is an extremely powerful tool to describe a vast range of physical phenomena in which CRs are involved. Below I discuss two of such applications, in their simplest form, while the same problems will be discussed in more detail in dedicated lectures. The first problem is that of describing the transport of CRs in the Galaxy (or any galaxy) for protons (the effect of nuclear spallation is easy to include but it will be done in a dedicated lecture, by C. Evoli). The second problem is that of diffusive particle acceleration at a non-relativistic shock front (Lecture Notes by D. Caprioli). Both these parts should not be considered as exhaustive descriptions, but rather as technical illustrations of how the main results of CR propagation and acceleration can be derived using the transport equation derived above. 

\subsection{Transport of CR protons in the Galaxy}
\label{sec:CRprop}
\begin{figure}
\centering
\includegraphics[width=1\textwidth]{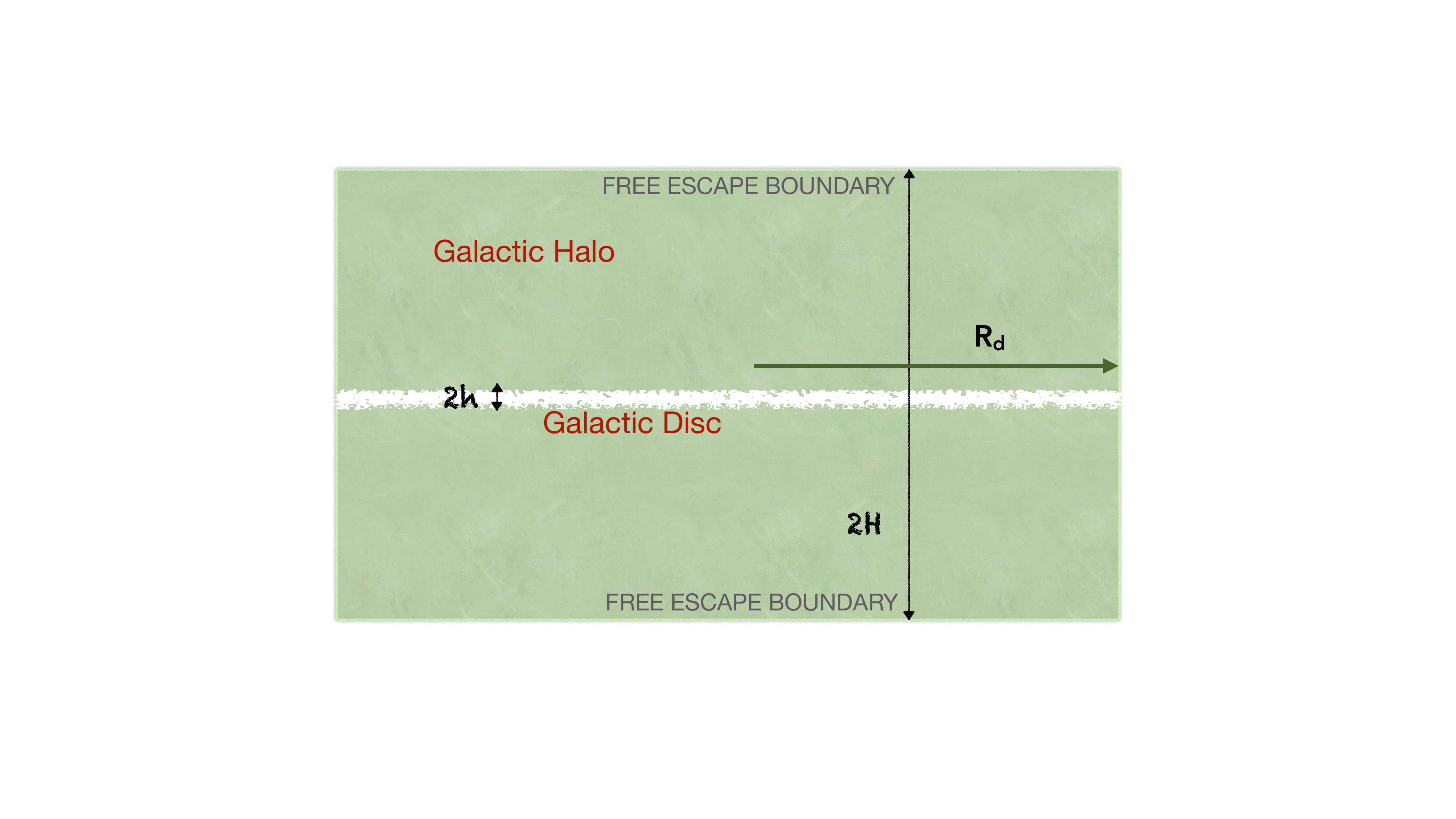}
\caption{Schematic view of the Galaxy comprising a thin disc and an extended halo.}
\label{fig:galaxy}
\end{figure}

Let us consider a simple model of the Galaxy as made of an infinitely thin disc where the sources are located, and an extended halo of half-thickness $H$ along the $\hat z$ direction, perpendicular to the plane of the disc, where CRs diffuse (see Fig. \ref{fig:galaxy}). I will refer to this extended region as the Galactic halo. Given the cylindrical symmetry of the problem, it is natural to assume that the problem is mainly sensitive to what happens in the $\hat z$ direction, which is also the direction in which CR gradients will be established. This assumption fails at locations sufficiently close to the edge of the disc, but here I will neglect such complications (see \cite{bible} for a discussion of these aspects). 

For simplicity I will assume here that the transport is purely diffusive, namely that there are no plasma motions, such as winds or generic outflows. Moreover, inspired by the estimates of the confinement time inferred from secondary/primary ratios and the abundance of unstable isotopes (see discussion in Sec. \ref{sec:how}), we know that such time is much shorter than the age of the Galaxy, so that it is reasonable to assume that a stationary state is reached. In this case the transport equation simply reads:
\begin{equation}
    \frac{\partial}{\partial z} \left[ D(p)\frac{\partial f}{\partial z}\right]= -Q(p,z),
    \label{eq:diff_galaxy}
\end{equation}
where $Q(p,z)$ represents a source term, namely the density of sources in phase space. Let us assume that the sources are in the form of impulsive events (namely with short duration compared with transport effects) occurring at a rate ${\cal R}$, each contributing a spectrum $N(p)$. Let also $\Sigma$ be the surface of the disc and the sources be located in an infinitely thin disc. In this case we can write $Q(p,z)=N(p){\cal R}\delta(z)/\Sigma$.
Clearly this simple equation, containing $\delta(z)$, expresses the fact that outside the disc the diffusive current, $J=-D\frac{\partial f}{\partial z}$, remains spatially constant. 

Eq. \ref{eq:diff_galaxy} can be easily solved once suitable boundary conditions are adopted at $z=0$ and $z=|H|$. At $z=0$ we require that the solution is the distribution function in the disc $f_d$, to be found as illustrated below. At the edge of the halo, $|z|=H$ we assume the so called free escape boundary condition, $f(p,|z|=H)=0$. From the physical point of view, requiring that $f$ vanishes at the boundary means that particles are no longer diffusing, hence their number density should drop. We will comment further on the implications of this assumption below. 

For $z>0$ (the solution is symmetric with respect to $z$), the condition that $D\frac{\partial f}{\partial z}$ vanishes implies that 
\begin{equation}
    f(p,z)=A+Bz ~~~~~~~ z>0,
\end{equation}
and imposing the two boundary conditions listed above, we obtain:
\begin{equation}
    f(p,z)=f_d(p)\left(1-\frac{|z|}{H}\right).
\end{equation}
This also means that $D\frac{\partial f}{\partial z}=-D\frac{f_d}{H}$ for ($z>0$). 

If we now integrate Eq. \ref{eq:diff_galaxy} in a narrow neighborhood of $z=0$ between immediately below and immediately above the disc, and we take into account the symmetry of the problem, we get:
\begin{equation}
    -2 D \frac{\partial f}{\partial z}|_{z=0^+}=\frac{N(p){\cal R}}{\Sigma},
\end{equation}
that, after using the derivative calculated above, becomes;
\begin{equation}
    f_d(p)=\frac{N(p){\cal R}}{2\Sigma}\frac{H}{D(p)} = \frac{N(p){\cal R}}{2\Sigma H}\frac{H^2}{D(p)}.
    \label{eq:fd}
\end{equation}
In the last step we simply multiplied and divided by $H$ in order to get the product of two quantities with an obvious physical meaning: the first fraction is the rate of injection per unit time averaged over the entire volume that can be filled with CRs, $V=2\Sigma H$. The second fraction is the diffusion time over a distance $H$. 
Eq. \ref{eq:fd} is then telling us that the equilibrium CR distribution in a point in the disc of the Galaxy is given by the product of the rate of injection per unit volume over the entire volume V and the confinement time of CRs in the same volume of the Galaxy. Notice that this result holds at any momentum $p$ of the particles: if we assume that the spectrum of an individual source is $N(p)\propto p^{-\alpha}$ and that the diffusion coefficient is also a power law function of momentum, $D(p)\propto p^\delta$, then the spectrum observed by an observer in the disc is not the source spectrum, but $f_d\propto p^{-\alpha-\delta}$.

This simple calculation illustrates in a very clear way one of the main features of Galactic CR: the spectrum of CRs that we observe is steeper than the one injected by the sources, as a consequence of the energy dependence of the diffusion coefficient. In slightly more refined versions of the same calculations (see \cite{Evoli2019,Evoli2020}) one can relate the momentum dependence of $D(p)$ to observed quantities, such as the ratio of fluxes of secondary and primary nuclei. This class of measurements confirms that $D(p)$ depends on momentum in a non trivial way. 

A few comments are in order concerning the meaning of the free escape boundary condition adopted above: in a more physical way we can assume that instead of being zero, the distribution function at $z=|H|$ is small, say $f_{*}(p)$. Independent of its exact value the flux of particles crossing the surface at $z=|H|$ must be conserved, hence it is clear that 
\begin{equation}
    -D\frac{\partial f}{\partial z}|_{z=H} = \frac{c}{2}f_*,
\end{equation}
where I assumed for simplicity that the particles move ballistically at $c$ outside the halo, since there is no diffusion, by assumption. The factor 2 accounts for the fact that the distribution of CRs outside the halo is isotropic on a hemisphere. Using the results above, this condition translates to 
\begin{equation}
    D\frac{f_d}{H} = \frac{c}{2}f_* \to f_*(p)=\frac{2D(p)}{c H} \approx \frac{2}{3}\frac{\lambda(p)}{H}f_d(p). 
\end{equation}
In the last step I replaced the diffusion coefficient with the diffusive pathlength following the general definition: $D(p)=(1/3) c \lambda(p)$. The condition that $f_*\ll f_d$ implies that the free escape boundary condition is a good approximation as long as the pathlength is much smaller than the size of the halo, $\lambda(p)\ll H$. Since $\lambda$ increases with the momentum $p$, the assumption will be best for low momenta. In practice however, for all energies of concern for Galactic CRs (say, below the knee), the free escape boundary condition works very well. 

The flux of particles escaping the Galaxy is then
\begin{equation}
    \phi_{esc}(p)=D(p)\frac{f_d}{H}=\frac{N(p){\cal R}}{2\Sigma}.
    \label{eq:escapeFlux}
\end{equation}
One should appreciate that the total spectrum escaping the Galaxy is in fact exactly the spectrum produced by the individual sources: in other words, if the observers were sitting outside the Galaxy, they would see the source spectrum and not the equilibrium spectrum that an Earth-bound observer is forced to measure. 

\subsection{Diffusive shock acceleration}
\label{sec:shock}

\begin{figure}
\centering
\includegraphics[width=1\textwidth]{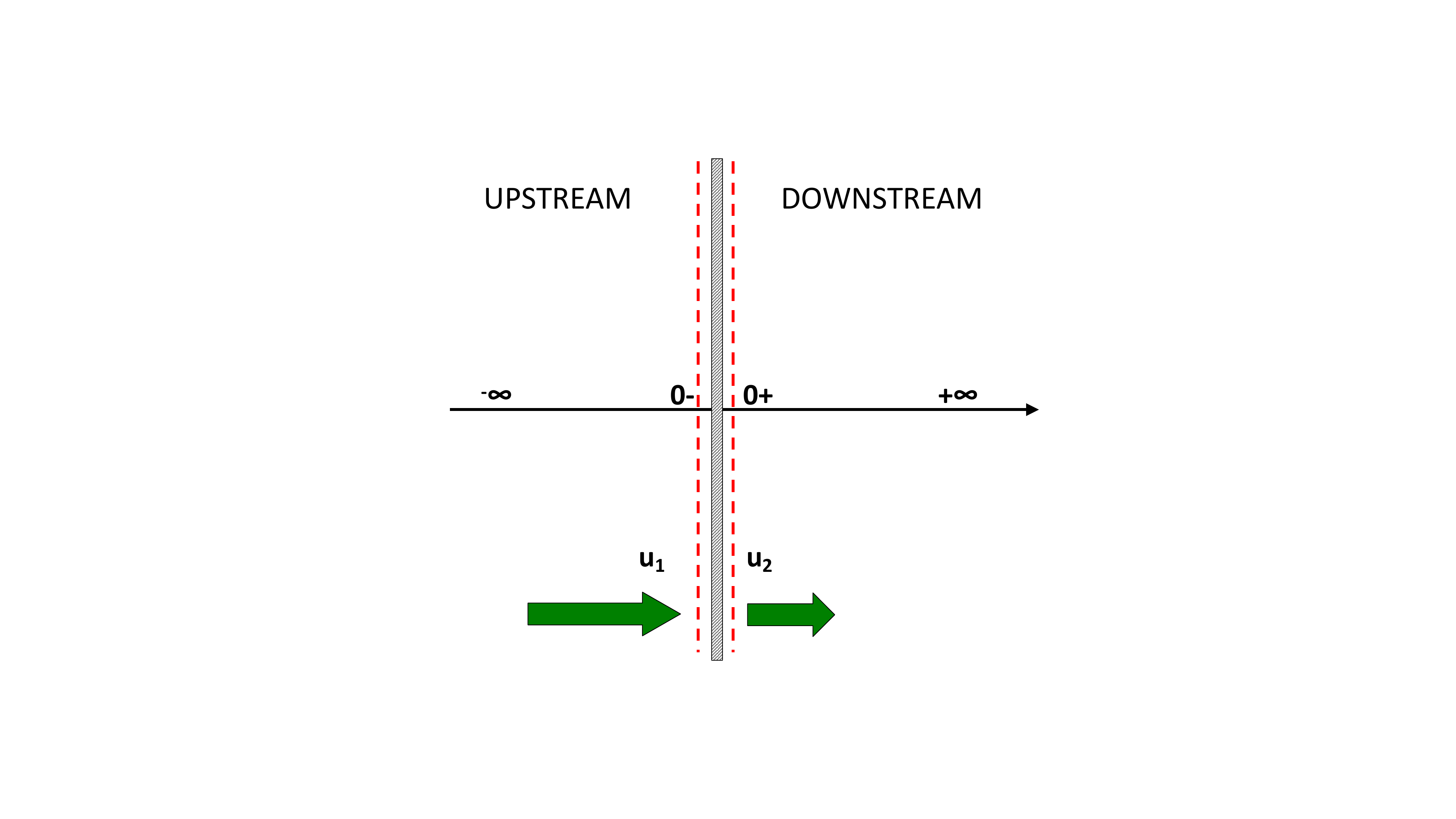}
\caption{Schematic view of a plane parallel shock.}
\label{fig:shock}
\end{figure}

In this section I will illustrate how the main results concerning the acceleration of charged particles at a non-relativistic shock may be easily obtained using the transport equation derived above. Let us consider a one-dimensional infinite shock front, propagating along the $\hat z$ direction. In the shock frame, the incoming plasma (upstream plasma) is moving towards the shock with velocity $u_1$ from $z=-\infty$. At the shock location, the gas is heated up and slowed down to a velocity $u_2$ (see Fig. \ref{fig:shock}), as inferred from the Rankine-Hugoniot relations (see for instance \cite{Vietri_book}). 
As it is usually the case, I introduce the compression factor $r=u_1/u_2$. Notice that the assumption of stationarity that I am adopting throughout this section is meaningful only in the shock frame. The shock is assumed to be infinitely thin, which is clearly just a mathematical description in that the physical thickness of a real shock is determined by the microphysics of the processes that are responsible for the formation of the shock front. In most astrophysical situations, shocks are collisionless, namely they are not formed because of physical collisions between particles but rather due to electromagnetic instabilities. Such processes are responsible for dissipation (heating) at the shock front and in a magnetized plasma they are expected to occur on scales of the order of the Larmor radius of the plasma particles. While this point is of little impact on the simple calculation presented here, it is of the utmost importance to understand which particles the transport equation applies to. In crossing the shock surface most of the particles in the plasma dissipate a large fraction of their kinetic energy into internal motion (the temperature increases in the downstream region). A few particles, the ones on the tail of the thermal distribution that happen to have a Larmor radius larger than the shock thickness and that are moving in the favorable direction, may cross the shock and bootstrap the acceleration process. The particles that are injected in this way are, by definition, not distributed isotropically. 

Since Eq. \ref{eq:transport_space_motion} refers to the isotropic part of the distribution function, it is clear that this equation cannot be applied to those particles whose momentum is  too close to the injection momentum. On the other hand, if the particle momentum is substantially larger than the injection momentum, then the particles are sensitive to the presence of the shock only because the plasma velocity is discontinuously changing, while the distribution function must be continuous across the shock. This is a mathematical way to impose that the accelerated particles described by the transport equation are not part of the thermal plasma. 

If the shock thickness is neglected, the derivative of the plasma velocity in the shock frame can be simply written as:
\begin{equation}
    \frac{du}{dz}=(u_2-u_1)\delta(z). 
\end{equation}
Moreover, let us assume that the shock structure is stationary (as pointed out above, this is a meaningful assumption only in the reference frame of the shock), so that the transport equation simplifies to:
\begin{equation}
    u\frac{\partial f}{\partial z} - \frac{1}{3} \frac{du}{dz} p \frac{\partial f}{\partial p}
    =\frac{\partial}{\partial z}\left[D\frac{\partial f}{\partial z}\right],
    \label{eq:transport_shock}
\end{equation}
where for simplicity I have used the notation $D(p)$ for the spatial diffusion coefficient, while I will allow the diffusion properties upstream and downstream of the shock to be different. 

Let us first consider separately the upstream and downstream sections. In each one of them the velocity of the plasma is spatially constant, namely $du/dz=0$). This means that, for instance, upstream of the shock:
\begin{equation}
    \frac{\partial}{\partial z}\left[ u f - D\frac{\partial f}{\partial z}\right] = 0 \to u f - D\frac{\partial f}{\partial z} = \rm constant,
    \label{eq:transport_up}
\end{equation}
and since at upstream infinity all quantities must vanish ($f(z=-\infty)=0$ and $\partial f/\partial z|_{z=-\infty}=0$), it follows that the constant is zero and that the solution must have the form:
\begin{equation}
    f(p,z)=f_s(p)\exp\left[ \frac{u_1 z}{D_1}\right],~~~~~~~z<0,
\end{equation}
where $f_s(p)$ is the distribution function evaluated at the location of the shock. The exponential shape of the solution reflects an important physical point: diffusion makes the particles preferentially move along their gradient, namely towards upstream infinity, while advection pushes the particles back toward the shock front. The balance between these two trends is at a distance from the shock $z_d=D_1/u_1$, the so called diffusion length. For a given momentum $p$ of the particles the diffusion length is independent of time. 

In principle one could repeat the same line of reasoning in the downstream region. However, downstream of the shock both diffusion and advection tend to take particles away from the shock surface. This means that if the spatial distribution of accelerated particles at given momentum had an exponential shape with some scale height, at a later time this point would have moved toward larger values of $z$, thereby contradicting the assumption of stationarity. It follows that the only solution, in the downstream region, that is compatible with stationarity is the one for which $f(p,z)=f_s(p)$, independent of $z$. 

This result implies that 
\begin{equation}
    D_1\frac{\partial f}{\partial z}|_{0^-}=u_1 f_s,~~~~~~~D_2\frac{\partial f}{\partial z}|_{0^+}=0.
\end{equation}
Let us now consider again Eq. \ref{eq:transport_shock} and integrate it in a narrow neighborhood of the shock, between $z=0^-$ and $z=0^+$, and recall that $f$ is continuous across the shock. This procedure results in:
\begin{equation}
    \frac{1}{3}(u_2-u_1)p\frac{d f_s}{d p}=u_1 f_s.
\end{equation}
This is a first order differential equation for $f_s(p)$ which has a simple power law solution
\begin{equation}
    f_s(p)=K p^{-\alpha},~~~~~~~\alpha=\frac{3u_1}{u_1-u_2}=\frac{3r}{r-1},
\end{equation}
where we used the definition of compression factor at the shock $r=u_1/u_2$. The constant $K$ is related to the efficiency of acceleration at the shock and will be discussed in the lectures of D. Caprioli. 

It is worth stressing that the power law shape that we just obtained is the trademark of non-thermal particles: we simply used the transport equation applied to suprathermal particles in a plasma with a sharp velocity gradient to infer that particles get energized in such a system, by repeatedly crossing the shock surface. We just derived diffusive shock acceleration as a straightforward application of the transport equation to a plane parallel shock, a further confirmation of the power of such an equation. 

A few comments are in order: 
\begin{itemize}
    \item The spectrum of accelerated particles is a power law, with a slope that only depends on the compression factor: the independence of the result on the diffusion coefficients, expression of the microphysics, is one of the reasons for the immense success of the theory in that its essence does not rely on the poorly known microphysics. 
    \item On the other hand, this independence is also one of the weak points of the theory: the stationary assumption that we adopted does not leave room to the existence of a maximum momentum (where the solution would be forced to depart from a power law, thereby not being solution of the transport equation). One way to impose by hand a maximum momentum is by requiring a free escape boundary condition upstream (or downstream) of the shock.  
    \item The microphysics of particle transport (namely the diffusion coefficients) is responsible for the definition of diffusion length and diffusion time. Together with the finite size or finite temporal extent of the shock, this leads to the determination of the maximum momentum of accelerated particles in a realistic setup.  
    \item For a strong shock (Mach number much larger than unity) the compression factor asymptotically tends to $r=4$, which makes the spectrum universal and $f_s(p)\propto p^{-4}$. Notice that the spectrum is a power law in momentum, not in energy: for relativistic particles $f_s(E)\propto E^{-2}$, while for non-relativistic particles $f_s(E)\propto E^{-3/2}$.
\end{itemize}
This simple theoretical approach (see \cite{Bell1978}) has been extremely useful to understand the basic physical effects involved in diffusive shock acceleration: particle acceleration occurs despite the absence of large scale electric fields, because the velocity gradient in the plasma velocity leads to an irreducible electric field, which in turn causes particle energization. Notice that when a particle crosses the shock from upstream to downstream or viceversa the electric field is always directed in such a way as to increase the energy of the particle. Physically, this is the reason why this acceleration process is firts order in the quantity $(u_1-u_2)/c$.

The same simple theoretical approach also contains the seeds to understand its limitations and even failures: first, notice that the $p^{-4}$ spectrum is energy divergent, in the sense that $\int_0^\infty f_s(p) p^2 p v(p)$ diverges logarithmically in the asymptotic case $r=4$. One could object, correctly, that this is only an artifact of the assumption of stationarity which does not allow a maximum momentum to exist. However, it remains true that in a simple theory like the one we discussed above, the energy carried by the accelerated particles can be larger than $\rho u_1^2$, the total ram pressure carried by the upstream plasma. Whether this is the case only depends on the normalization constant in front of the spectrum, that cannot be calculated in this simple approach. In more detailed calculations, as well as in numerical simulations, this actually happens. It is clear that a non-linear theory must include the dynamical effects of the accelerated particles on the background plasma. An in-depth discussion of such a theory can be found in \cite{Malkov2001,Blasi2013}, while for a discussion of numerical simulations see \cite{Caprioli2014a,Caprioli2014b,Caprioli2014c,Caprioli2015} and Lecture Notes by D. Caprioli. 

Second, it has been noted since the beginning of the development of the theory of diffusive shock acceleration, that if to take the diffusion coefficient in the Galaxy as a proxy for the upstream diffusion coefficient, the maximum energy that one can estimate for galactic sources such as SNRs is exceedingly low (in the few GeV range). Hence particles themselves are required to produce their own perturbations, thereby reducing diffusivity, if to consider shocks as plausible sites of particle acceleration \cite{Lagage1983-1,Lagage1983-2,Bell2004} (see also Lectures by A. Marcowith). 

Both the dynamical reaction of accelerated particles and the self-generation of perturbations are required ingredients of a non-linear theory of particle transport, which will be discussed only in passing in Sec. \ref{sec:nonlin}. 

\section{An introduction to diffusion perpendicular to magnetic field lines}
\label{sec:perp}

The previous analysis of the problem of particle transport in a magnetized plasma referred to the motion of the particles parallel to the ordered magnetic field $\vec B_0$. However, particles can also propagate across field lines, mainly as a result of the so-called random walk of magnetic field lines. The diffusive nature of such motion is all but guaranteed and in general what happens in the perpendicular direction is not independent upon parallel transport. 

While the theory of parallel transport is well established, at least for weak perturbations, $\delta B/B_0\ll 1$, the theoretical approach to perpendicular transport is still subject of intense investigation. In these Lectures I will limit myself with laying down the main ingredients of the problem and list some of the seminal work that has been done to provide a description of perpendicular transport, while not adventuring in technical details. The main reason for this choice is that there is no consensus yet on such details, and it would be impossible to make justice to the plethora of approaches to this problem in the short space allocated for these Notes. As an instance, I stress that none of the current theoretical descriptions of perpendicular transport is able to account for the different energy dependence of the parallel and perpendicular diffusion coefficients in the case of isotropic turbulence \cite{Dundovic2020} on top of a regular magnetic field (see however the recent work in Ref. \cite{Mertsch23}). On the other hand, an exhaustive discussion of the theoretical methods put forward to describe this problem can be found in \cite{Shalchi_book}.

The basis of the problem of perpendicular transport follows an approach that can be understood using the tools we already built in Sec. \ref{sec:pitch}. Let us define a velocity in the $x$ or $y$ directions (perpendicular to the direction of $\vec B_0$) as $v_i$, with $i=\left\{x,y\right\}$. In a similar way we introduce the displacements in the $x$ and $y$ directions as $\Delta x_i$ with $(\Delta x_1,\Delta x_2)=(\Delta x,\Delta y)$ and
\begin{equation}
    \Delta x_i (t) = \int_0^t d\tau v_i(\tau).
\end{equation}
Following the same line of reasoning as for the parallel motion, we introduce the following second order correlator of spatial displacements:
\begin{equation}
    \langle \Delta x_i \Delta x_i \rangle = \langle \int_0^t d\tau v_i (\tau) \int_0^t d\xi v_i (\xi)\rangle = \int_0^t d\tau \int_0^t d\xi \langle v_i (\tau) v_i (\xi)\rangle.
\end{equation}
Splitting the second integral in two parts one can easily write:
\begin{equation}
    \langle \Delta x_i \Delta x_i \rangle = \int_0^t d\tau \int_0^\tau d\xi \langle v_i (\tau) v_i (\xi)\rangle + \int_0^t d\tau \int_\tau^t d\xi \langle v_i (\tau) v_i (\xi)\rangle.
\label{eq:deltaX2}
\end{equation}
It is useful to remark that if the perturbations are homogeneous in space, the correlator can only depend upon the time difference and not upon the absolute times. Hence:
\begin{eqnarray}
    \langle v_i (\tau) v_i (\xi)\rangle = \langle v_i (\tau-\xi) v_i (0)\rangle,~~~~~~~\rm if ~\tau>\xi \\
    \langle v_i (\tau) v_i (\xi)\rangle = \langle v_i (\xi-\tau) v_i (0)\rangle,~~~~~~~\rm if ~\xi>\tau. 
\end{eqnarray}
This allows us to rewrite Eq. \ref{eq:deltaX2} as:
\begin{equation}
    \langle \Delta x_i \Delta x_i \rangle = \int_0^t d\tau \int_0^\tau d\xi \langle v_i (\tau-\xi) v_i (0)\rangle + \int_0^t d\tau \int_\tau^t d\xi \langle v_i (\xi-\tau) v_i (0)\rangle,
\end{equation}
and changing variables to $\xi'=\tau-\xi$ in the first term and to $\xi'=\xi-\tau$ in the second:
\begin{equation}
    \langle \Delta x_i \Delta x_i \rangle = \int_0^t d\tau \int_0^\tau d\xi' \langle v_i (\xi') v_i (0)\rangle + \int_0^t d\tau \int_0^{t-\tau} d\xi' \langle v_i (\xi') v_i (0)\rangle.
\label{eq:deltaX2_1}
\end{equation}
If we multiply by $d\tau/d\tau$ in the integral in $\tau$ and integrate by parts we obtain:
\begin{eqnarray}
    \langle \Delta x_i \Delta x_i \rangle = \int_0^t d\tau \frac{d\tau}{d\tau}\int_0^\tau d\xi' \langle v_i (\xi') v_i (0)\rangle + \int_0^t d\tau \frac{d\tau}{d\tau}\int_0^{t-\tau} d\xi' \langle v_i (\xi') v_i (0)\rangle =\nonumber \\
    \int_0^t d\tau (t-\tau) \langle v_i (\tau) v_i (0)\rangle + \int_0^t d\tau \tau \langle v_i (t-\tau) v_i (0)\rangle \nonumber,
\end{eqnarray}
and changing variable in the second integral, $\tau'=t-\tau$:
\begin{equation}
    \langle \Delta x_i \Delta x_i \rangle = 2\int_0^t (t-\tau) \langle v_i(\tau) v_i (0)\rangle.
\end{equation}
At this point we can introduce the so-called running spatial diffusion coefficient as 
\begin{equation}
    {\cal D}_{ii} = \frac{1}{2} \frac{d}{dt} \langle \Delta x_i \Delta x_i \rangle = \int_0^t d\tau \langle v_i(\tau) v_i (0)\rangle.
\end{equation}
If the motion of the particles is diffusive, in the limit of large time $t$ this quantity becomes constant and plays the role of spatial diffusion coefficient:
\begin{equation}
    D_{ii}=\int_0^\infty d\tau \langle v_i(\tau) v_i (0)\rangle.
    \label{eq:TGZ}
\end{equation}

In fact, the formalism introduced above, known as the Green-Kubo-Taylor formalism \cite{Taylor1922,Green1954,Kubo1957}, applies equally well to both parallel ($i=z$) and perpendicular ($i=x,y$) transport. For parallel motion $v_z(t)=v\mu(t)$, as I discussed in Sec. \ref{sec:pitch}.

In order to apply this formalism to the problem of perpendicular transport I need to express $v_{x,y}(t)$ as functions of the perturbations in the magnetic field. With this in mind, let us reconsider the equations of motion in Eqs. \ref{eq:mot1},\ref{eq:mot2} but retaining the first order correction in the small perturbations:
\begin{eqnarray}
    \frac{d v_x}{dt} = \Omega \left(v_y - v_z\frac{\delta B_y}{B_0} \right)\\
    \frac{d v_y}{dt} = \Omega \left(-v_x + v_z\frac{\delta B_x}{B_0} \right).
    \label{eq:vel}
\end{eqnarray}
In the case in which the perturbative term is neglected, the solution is in the form of a sum of cosine and sine terms, with constants chosen to satisfy the initial conditions. In the more general case considered here, let us look for a solution in the following form:
\begin{equation}
    v_x(t)=A(t)\cos(\Omega t)+B(t)\sin(\Omega t).
\end{equation}
Inspired by the analogy with the case in which we neglect the perturbative term, I choose the following functional form for $v_y$:
\begin{equation}
    v_y(t)=B(t)\cos(\Omega t)-A(t)\sin(\Omega t).
\end{equation}
Substituting $v_x$ and $v_y$ in Eqs. \ref{eq:vel}, I find that the following identities must hold:
\begin{eqnarray}
    \dot A \cos(\Omega t) + \dot B \sin(\Omega t) = -\Omega v_z \frac{\delta B_y}{B_0}\equiv - \alpha(t) \\
    \dot B \cos(\Omega t) - \dot A \sin(\Omega t) = \Omega v_z \frac{\delta B_x}{B_0}\equiv  \beta(t).
\end{eqnarray}
These algebraic equations can be readily solved to give:
\begin{eqnarray}
    \dot B(t) = -\alpha(t) \sin(\Omega t) + \beta(t) \cos(\Omega t) \\
    \dot A(t) = -\alpha(t) \cos(\Omega t) - \beta(t) \sin(\Omega t). 
\end{eqnarray}
At this point a simple integration in time returns the two functions $A(t)$ and $B(t)$:
\begin{eqnarray}
    A(t)=v_{x,0}-\int_0^t dt' \left[ \alpha(t') \cos(\Omega t') + \beta(t') \sin(\Omega t')\right],\nonumber\\
    B(t)=v_{y,0}+\int_0^t dt' \left[ \beta(t') \cos(\Omega t') - \alpha(t') \sin(\Omega t')\right],
\end{eqnarray}
where I used the initial conditions $A(t=0)=v_{x,0}$ and $B(t=0)=v_{y,0}$. 

Based on our findings, the particle velocity in the $x$ direction can be written as:
\begin{eqnarray}
    v_x(t) = v_{x,0}\cos(\Omega t) + v_{y,0}\sin(\Omega t) - \nonumber \\
    -\cos(\Omega t) \int_0^t dt' \left[ \alpha(t') \cos(\Omega t') + \beta(t') \sin(\Omega t')\right] + \nonumber \\
    +\sin (\Omega t) \int_0^t dt' \left[ \beta(t') \cos(\Omega t') - \alpha(t') \sin(\Omega t')\right].
    \label{eq:vx}
\end{eqnarray}
A similar expression can be written for $v_y(t)$.

Notice that the functions $\alpha(t)$ and $\beta(t)$ calculated at time $t$ should contain the value of the perturbed field $\delta B_{x,y}$ at the location of the particle at time $t$, which, given the complex motion of the particle, is not known {\it a priori}. 

Most approaches to the description of perpendicular transport make the assumption that the guiding center of the particles follows the magnetic field line: this means that in fact we are not interested exactly where the particles are at time $t$ along the approximately circular gyro-orbit, but rather where the guiding center is. Let us introduce the velocity of the guiding center as:
\begin{equation}
    \tilde v_{x,y}(t) = \frac{1}{T} \int_t^{t+T} dt' v_{x,y} (t'),
    \label{eq:vxtilde}
\end{equation}
where $T=2\pi/\Omega$ is the gyroperiod. The first two terms in Eq. \ref{eq:vx} clearly give no contribution to the guiding center velocity because of the periodicity of the cosine and sine functions over one gyroperiod. 

At this point I make one additional assumption that is however rather well justified, at least for most situations of interest, namely that the perturbations $\delta B_{x,y}$ are slowly varying functions of time over one gyroperiod. This means that in one gyration the particles experience the same value of $\delta B_{x,y}$. With this last assumption, the functions $\alpha$ and $\beta$ can be taken out of the integrals and it is straightforward to see that 
\begin{equation}
    \tilde v_x(t) \approx v_z \frac{\delta B_x}{B_0},
\end{equation}
and similarly
\begin{equation}
    \tilde v_y(t) \approx v_z \frac{\delta B_y}{B_0}.
\end{equation}
If I also assume that the guiding center velocity in the $\hat z$ direction equals $v_z(t)$, then we must conclude that, as long as the perturbations change slowly in time, the guiding center of the motion of a particle must follow the magnetic field lines. 

At this point, the spatial displacement of the particles in the direction perpendicular to $\vec B_0$ can be interpreted as Eq. \ref{eq:TGZ} applied to the guiding center. For instance:
\begin{equation}
D_{xx} = \int_0^\infty d\tau \langle \tilde v_x(\tau) \tilde v_x(0)\rangle = \left( \frac{1}{B_0^2}\right)^2 \int_0^\infty d\tau \langle v_z(\tau) v_z(0) \delta B_x(\tau) \delta B_x(0)\rangle.
\label{eq:Dxx}
\end{equation}
Clearly in normal conditions $D_{xx}=D_{yy}$. Eq. \ref{eq:Dxx} clearly illustrates the difficulties in evaluating the diffusion coefficient perpendicular to $\vec B_0$: mathematically it results in a fourth order correlator which in general is hard to determine, unless some simplifying assumptions are made. 

Different theoretical models of perpendicular transport rely on different assumptions on how to treat this fourth order correlation. An exhaustive treatment of the topic can be found in \cite{Shalchi_book}. A particularly useful approach is called non-linear guiding center theory (NLGCT) \cite{Matthaeus2003}, in which several assumptions are made: first, it is assumed that the fourth order correlation can be split in the product of two second order correlation functions, $\langle v_z(\tau) v_z(0) \delta B_x(\tau) \delta B_x(0)\rangle = \langle v_z(\tau) v_z(0)\rangle \langle \delta B_x(\tau) \delta B_x(0)\rangle$. Second, an ansatz is adopted for the correlator of $v_z$ (for instance random ballistic approximation or parallel diffusion with an assigned parallel pathlength); third, an ansatz is introduced to allow one to write the magnetic correlation function in terms of quantities that we can handle (for instance the power spectrum of the perturbations). This procedure, in which the second and third steps may differ, leads to testable predictions for the perpendicular diffusion coefficient as a function of the parallel mean free path (the latter must be known {\it a priori}, it is not an output of NLGCT). For simple models of the turbulence, for instance slab+2D models, the NLGCT has provided useful insights. However, for substantial levels of perpendicular complexity and for turbulence that cannot be separated in slab+2D components, the NLGCT fails to reproduce findings of numerical simulations (see for instance \cite{Dundovic2020}).

\section{Non linear theories of parallel particle transport}
\label{sec:nonlin}

The diffusive motion that we have described in the previous sections is due to the interaction of the charged particles with the microscopic random fields associated with an ensemble of perturbations. The net effect of these interactions is to isotropize the directions of motion of the particles in the reference frame of the waves, or, in other words, to slow down the bulk motion of the particles to the same speed as the waves. 

This interpretation, in its simplicity, leads to a rather immediate implication: imagine to have a beam of charged particles moving with a drift (bulk) speed $v_D$ in, for instance, the $\hat z$ direction. After the process of isotropization, the bulk velocity of the particles is reduced (in the Lab frame) to about the Alfv\'en speed. This means that there has been a net loss of momentum in the $\hat z$ direction. Such momentum is transferred to the background plasma in the form of heat, bulk motion and creation of other perturbations. This last component appears in the form of magnetic field amplification, and is of the utmost importance in the proximity of acceleration regions, but also for the description of CR transport in the Galaxy and around sources. 

A formal derivation of this result can be obtained by studying the dispersion relation of waves in the presence of non thermal particles (see Lectures by A. Marcowith), but an order of magnitude of the effect can be obtained following a simple argument proposed by R. Kulsrud \cite{Kulsrud_book}.

Assuming that the drift velocity of the cloud of particles in the $\hat z$ direction is $v_D\ll c$, one can say that the momentum in the same direction is $n_{CR}m_p v_D$ and the change occurs over a time that is of the same order of magnitude as the time required for deflection by 90$^o$, namely
\begin{equation}
\tau_{rev} \sim \frac{1}{D_{\theta\theta}} \approx \frac{1}{\Omega {\cal F}(k_{res})}
\end{equation}
where $k_{res}$ is the wavenumber resonant with the gyration of particles of momentum $p$ dominating the current of CR moving with drift velocity $v_D$. The rate of momentum change is then
\begin{equation}
    \frac{dP_{CR}}{dt} = n_{CR} m_p (v_D-v_A) \Omega {\cal F}.
\end{equation}
This momentum change is compensated by an equal momentum increase of the waves. If we introduce the rate at which this phenomenon occurs, $\Gamma_W$, we can write:
\begin{equation}
     \frac{dP_{w}}{dt} = \Gamma_{w}\frac{\delta B^2}{8\pi} \frac{1}{v_A}.
\end{equation}
Equating these two expressions leads to an estimate of the growth rate that, neglecting factors of order unity, reads:
\begin{equation}
    \Gamma_w \approx \frac{n_{CR}(>p)}{n_i}\frac{v_D-v_A}{v_A} \Omega_{cyc},
    \label{eq:gammaW}
\end{equation}
where $\Omega_{cyc}=q B_0/m_p c$ is the cyclotron frequency. In the last expression I wrote more explicitly that the CR density refers to particles with momentum $>p$, so that the $k=1/r_L(p)$ resonant wavenumber refers to the same particles. In terms of the distribution function in phase space we can write $n_{CR}(>p)\approx 4\pi p^3 f(p)$. 

This expression coincides with the one derived in a formal way within factors of order unity. The formal analysis proceeds through the solution of the dispersion relation of perturbations allowed in a plasma in the presence of non-thermal particles. When $v_D>v_A$ such an equation admits a complex solution with a positive imaginary part of the frequency, which thereby implies the excitation of an instability, known as resonant streaming instability \cite{Kulsrud-Pearce1969,Skilling1975streaming}. 

The expression in Eq. \ref{eq:gammaW} allows us to estimate the importance of streaming instability in a variety of situations, ranging from acceleration sites to propagation in the Galaxy. In the following I will briefly discuss a few instances, in order to illustrate the prominent role played by this phenomenon in modern cosmic ray astrophysics. Below, I will also occasionally mention another branch of the straaming instability, the non-resonant one \cite{Bell2004}, which has become increasingly more popular, for excellent reasons, since it was originally proposed. 

\subsection{Self-generation in DSA}
\label{sec:SG_DSA}

As we discussed earlier in these Lecture Notes, particle acceleration can take place in the proximity of a shock front and if the shock is strong (Mach number much larger than unity) the spectrum of the accelerated particles is $f(p)\propto p^{-4}$. If we assume that the pressure in the form of accelerated particles with momentum $p>p_0$, $P_{CR}=\frac{1}{3} \int_{p_0}^{p_{max}} dp 4\pi p^3 v f(p)$, is normalized in such a way that it is a fraction $\xi_{CR}$ of the ram pressure $\rho v_s^2$, we can write
\begin{equation}
    f(p)=\frac{3\xi_{CR}\rho v_s^2}{4\pi p_0^4 \Lambda} \left( \frac{p}{p_0}\right)^{-4},
\end{equation}
where $\Lambda=\ln\left(\frac{p_{max}}{p_0}\right)$, $\rho$ is the density of gas ahead of the shock and $v_s$ is the shock speed.

Replacing this result in Eq. \ref{eq:gammaW}, with $n_{CR}(>p)\approx 4\pi p^3 f(p)$, I obtain:
\begin{equation}
    \Gamma_W = \frac{3\xi_{CR}\rho v_s^2 }{c \Lambda p} \frac{1}{n_i}\frac{v_s}{v_A}\frac{q B_0}{m_p c} =
    \frac{3\xi_{CR}}{\Lambda} M_A^2 \frac{v_s}{c} \frac{v_A}{r_L(p)},
\end{equation}
where I introduced the Alfv\'enic Mach Number $M_A=v_s/v_A$. If one assumes the canonical reference value of $\xi_{CR}=10\%$ for the CR acceleration efficiency, it is easy to see that the characteristic time of growth of the streaming instability is $\Gamma_W^{-1}\sim 3\times 10^5$ s for particles of momentum $p=1$ GeV/c and $\Gamma_W^{-1}\sim 10^4$ years for particles of momentum $p=10^6$ GeV/c.

This simple estimate shows in a rather clear way why it is so difficult to accelerate particles to $\sim \rm PeV$ energies in SNR shocks: the process is effective only if waves are efficiently produced due to the same accelerated particles. The previous estimates shows that the growth time associated to particles of $PeV$ energies is much longer that the Sedov time, where most energetic particles are accelerated. Hence the excitation of the resonant streaming instability is not sufficient to accelerate particles to PeV energies in SNR, as was already understood in \cite{Lagage1983-1,Lagage1983-2}.

As recently discussed by \cite{Cristofari2020,Cristofari2021}, even the excitation of the non-resonant hybrid instability, that typically grows much faster, requires rather extreme conditions for the acceleration to PeV energies, conditions that are likely to be realized only in very energetic rare supernova events. 

\subsection{Self-generation in Galactic CR transport}
\label{sec:SG_Galaxy}

An inspection of the transport equation in its stationary form immediately shows that the quantity $(v_D-v_A)n_{CR}(>p)$ is related to $D 4\pi p^3\frac{\partial f}{\partial p}$. In its simplest form, as discussed above, the transport equation for protons in the Galaxy leads to $\partial f/\partial z=f_d/H$, hence the growth rate in Eq. \ref{eq:gammaW} can be rewritten as:
\begin{equation}
    \Gamma_W (k) \approx \frac{n_{CR}(>p)}{n_i} \frac{v_D-v_A}{v_A}\Omega_c \approx \frac{1}{n_i} \frac{\sqrt{4\pi \rho}}{B_0}\frac{q B_0}{m_p c} D\frac{\partial f}{\partial z} 4\pi p^3 \approx \frac{P_{CR}(>p)}{U_B}\frac{v_A}{H} \frac{1}{{\cal F}(k)},
\end{equation}
where $k=1/r_L(p)$ is the resonant wavenumber for particles with momentum $p$. For the CRs that carry most of the energy on Galactic scales, $P_{CR}/U_B\sim 1$ and, as we discussed in Sec. \ref{sec:pheno}, ${\cal F}\sim 10^{-5}$ for such particles. Hence, purely based upon the observed $B/C$ and $Be/B$ ratios, we can conclude that the resonant streaming instability grows on time scales $\Gamma_W\sim 500$ years, much shorter than the confinement time of Galactic CRs. It follows that self-generation has to play an important role for Galactic transport, at least at relatively low energies. 

If the scattering centers responsible for CR transport in the Galaxy are completely self-generated, one can estimate the energy dependence of the diffusion coefficient: if most of the volume where CRs diffuse is completely ionized (halo) then the main reason why the instability stops growing is non-linear Landau damping, that occurs at a rate \cite{LeeVolk1973}: $\Gamma_{NLL}(k)=k c_s {\cal F}(k)$. Assuming stationarity, the equilibrium value of ${\cal F}$ can be calculated by imposing balance between growth and damping:
\begin{equation}
    \Gamma_W=\Gamma_{NLL} \to {\cal F}(k)\approx \left[ \frac{v_A}{H} \frac{r_L(E)}{c_s} \left(\frac{E}{GeV}\right)^{-0.7}\right]^{1/2} \sim 10^{-5} \left(\frac{E}{GeV}\right)^{0.15}, 
\end{equation}
where we assumed that $P_{CR}(>p)$ roughly scales as $E^{-0.7}$ and that the CR pressure and the magnetic pressure are roughly equal for particles in the GeV range. Using the quasi-linear expression for the diffusion coefficient as derived in Sec. \ref{sec:pheno} we easily obtain that $D(E)=\frac{1}{3}r_L v/{\cal F}\sim E^{0.85}$ for relativistic energies. 

A few comments are in order: first, we notice that in the context of non linear approaches to CR transport, the diffusion coefficient is an output of the calculations. This result is solely determined by the balance of growth and damping of waves. The second point to notice is that the balance between growth and damping, as estimated above, leads to a value of ${\cal F}$ for $\sim GeV$ particles that is of the correct order of magnitude necessary to account for Galactic CR confinement (see discussion in Sec. \ref{sec:pheno}). This result relies uniquely upon the measured spectrum of Galactic CR at the Earth, and it is therefore remarkable that the phenomenon of self-generation may warrant CR self-confinement, at least at low energies. 

Third point to notice is that typically the energy dependence of the self-generated diffusion coefficient is rather strong. In terms of diffusion in the Galaxy, this translates to a steep decrease in the confinement time of CRs, which immediately leads to the conclusion that self-confinement can only be effective at low energies, while at high energies the diffusive transport must be dominated by scattering off pre-existing turbulence (the so-called extrinsic turbulence). By the same token, the secondary/primary ratios are expected to drop rather quickly with energy and then harden when scattering becomes dominated by extrinsic turbulence.  

Some attempts have been made to go beyond this simple picture and interpret the halo of the Galaxy as an emerging phenomenon: in particular \cite{Evoli2018,Dogiel2021,Dogiel2022} proposed that the halo may be due to a combination of two phenomena, the self-generation of perturbations due to CRs themselves, and the advection of cascading turbulence while moving outwards away from the Galactic disc. These models are still subject of active investigation, due to the complex interplay of wave generation, damping and cascading. 

Following the line of reasoning inspired by non-linear self-generation of perturbations, at least qualitatively we are led to conclude that the spectra of protons, nuclei and even secondary products of spallation are expected to all have a break at the energy where diffusion from self-generated becomes dominated by scattering off extrinsic turbulence. More detailed calculations of this phenomenon \cite{Blasi2012,AloBla2013,AloBlaSer2015} suggest that the transition is likely to occur in the few hundred GeV range, and that it provides a possible explanation of the spectral breaks observed by PAMELA \cite{Pam-hard} and AMS-02 \cite{AMS-hard-p,AMS-hard-He} and recently confirmed by DAMPE \cite{DAMPE-hard} and CALET \cite{CALET-hard} (see \cite{Kempski2022} for a discussion of caveats of this picture). In this approach the break is expected to occur also in the secondary/primary ratios, since the transition affects directly the effective diffusion coefficient of particles. 

\subsection{Transport of CRs escaping sources}
\label{sec:escape}

The description of the transport of non-thermal particles outside their sources is one of the current hot research topics, for a variety of reasons, some of theoretical nature, others of observational nature. From the theoretical point of view, it is worth stressing that the spectrum of particles accelerated at a shock front, for instance in a supernova remnant, is, in general, different from the spectrum of particles released into the ISM. The connection between the escape spectrum and the spectrum of accelerated particles is complex and not fully understood. Such connection heavily relies upon the excitation of instabilities due to the particles leaking out of the acceleration region. From the observational point of view, the recent detection of regions of extended gamma ray emission around PWNe, the so-called TeV halos, has prompted considerable attention on the possibility that the small diffusion coefficients derived around such sources may be due to the same escaping particles, if they carry enough current. 

Close to sources of non-thermal particles we can reasonably expect strong gradients to be established: in the immediate vicinity of the source the density of such particles is the highest, while it eventually drops to the ambient mean density far from the source. The gradients imply the existence of currents, that may excite electromagnetic instabilities, if certain conditions are fulfilled. In turn, the random fields that are generated through these instabilities feed back onto the particles by slowing down diffusion, which in turn enhances the particle density in the source proximity. This chain of events gets eventually quenched when either the current is destroyed or the dynamical feedback on the background plasma becomes too large (for instance the plasma starts moving or bubbles are excavated inside the plasma, so as to reduce the pressure gradients). 

There are basically three situations in which these effects have been investigated: 1) escape of CRs from SNRs; 2) escape of CR from the Galaxy; 3) Escape of CRs from extragalactic sources of UHECRs. 

The effects of the particles accelerated at SNR shocks and escaping into the ISM was investigated by \cite{Ptuskin2008,Malkov2013,Nava2016,Nava2019,Recchia2022,DAngelo2016,DAngelo2018} with different assumptions on the conditions of the ISM in the circum-source region (ionization level, density, temperature). These assumptions mainly affect the damping of waves self-generated through particle transport in a complex chain of non linear effects, described by using non linear partial differential equations for the particles and the waves, coupled to each other through the self-generated diffusion coefficient, assumed to be well described by quasi-linear theory. These approaches are most reliable when the perturbations induced by particle transport are weak, so that the conclusions of quasi-linear theory can be trusted. As we discuss below, this is not always the case. 

The main effect of the excitation of the resonant streaming instability around SNRs is that the time that particles with energy $E\gtrsim$ TeV spend around the source is much longer than expected based on the Galactic diffusion coefficient. As a consequence, since most SNRs are located in the disc of the Galaxy where the density may be relatively high, the grammage accumulated around the source may become an appreciable fraction of the total grammage, as measured through B/C and similar ratios. However, when the gas density around the source is relatively high (of order $1~\rm cm^{-3}$), typically the medium is only partially ionized and this makes ion-neutral damping to have a prominent effect in limiting self-confinement. The increase in the confinement time in the circum-source region is most prominent for low gas density and high level of ionization, where the growth is only limited by non linear Landau damping. In this case the accumulated grammage is negligible, unless the source happens to be located in the proximity of a dense cloud, small enough to not affect appreciably wave growth, but dense enough to impact the grammage accumulated by particles. These effects have only been investigated in passing so far, but may be potentially important for a proper description of gamma ray observations.

\begin{figure}
\centering
\includegraphics[width=1\textwidth]{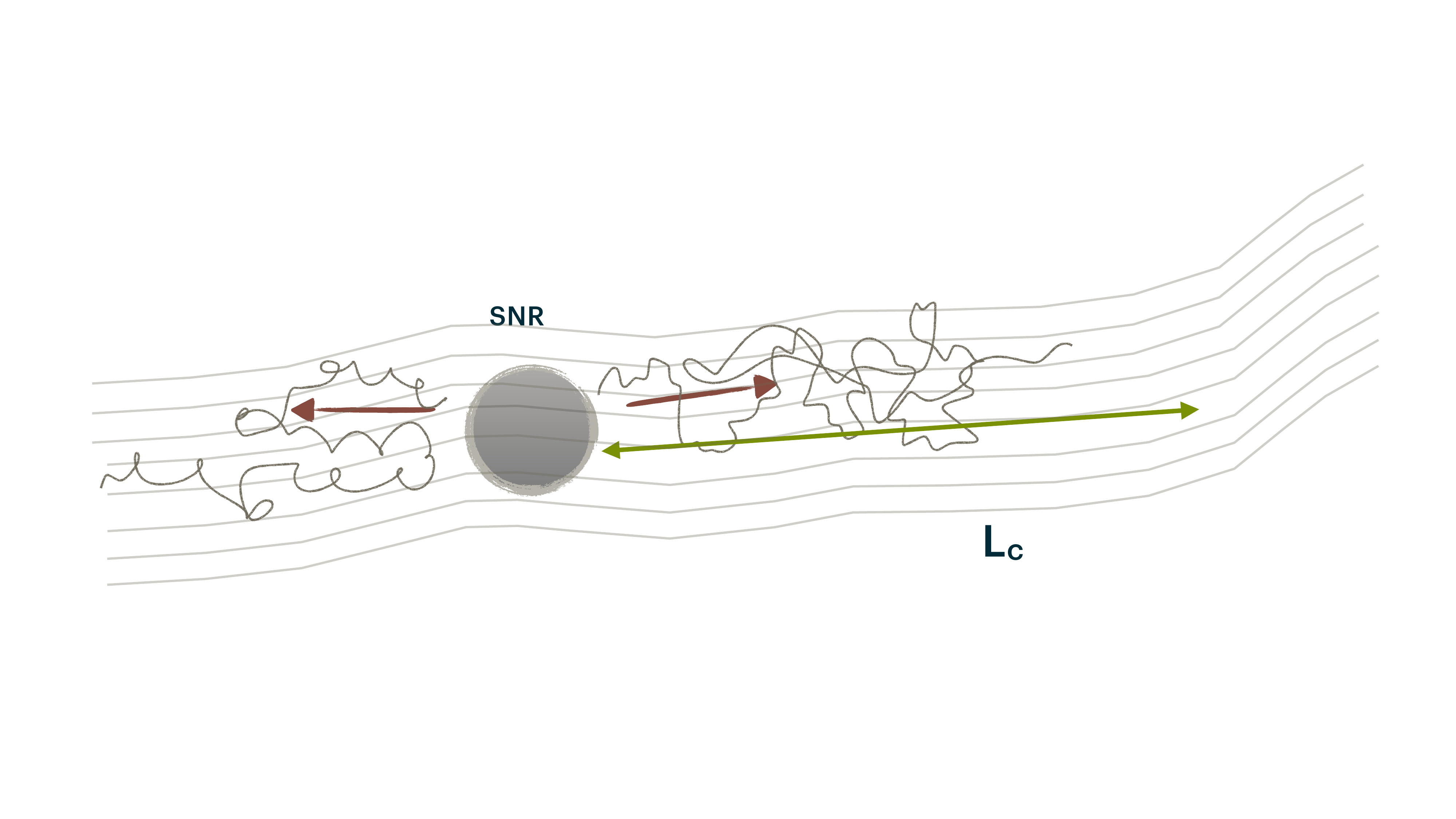}
\caption{}
\label{fig:fluxtube}
\end{figure}

In any case, it is not expected that self-generation may affect the transport of CRs with energy $E\gtrsim$ TeV, because the density of such particles is too low to lead to fast growth of the resonant modes. All these considerations apply to the simple case in which the {\it flux tube} approximation is made (see illustration in Fig. \ref{fig:fluxtube}): the particles are assumed to follow the magnetic field lines of the Galactic local magnetic field around the source. At first sight, the assumption appears to be justified since diffusion in the direction perpendicular to the magnetic field lines is expected to be much slower than diffusion in the parallel direction. Hence particles are expected to travel a distance of order the coherence scale of the Galactic magnetic field, $L_c$, faster than they can propagate perpendicular to the field lines. Once a distance of order $L_c$ is covered, the particles start moving in a magnetic environment that is more three dimensional, and the density of particles drops rapidly, reflecting the change of geometry from flux tube to 3D.  
However, a deeper look at the situation shows the existence of several subtleties that require special attention: in normal conditions (namely away from sources) the energy density of the ISM and of CRs is inferred to be in rough equipartition. This means that near sources, where the CR density is larger, especially if non-linear effects (self-generation) are at work, the CR energy density is bound to exceed the pressure of the background ISM plasma. Hence, even though in the beginning the flux-tube assumption may be valid, the CR energy density accumulated in the tube quickly becomes much larger than outside the tube. This implies that a strong pressure gradient is established in the direction perpendicular to the tube length, leading to the inflation of the tube. 

The second point to take into account is that for sufficiently high energy particles, the mean free path for parallel diffusion may be larger than $L_c$ (as a simple exercise one can estimate the pathlength for diffusion using the Galactic diffusion coefficient for particles of TeV energy and compare it with $L_c\sim 10-50$ pc). For such particles the motion starts as quasi-ballistic, so than an effective current develops and takes particles away from the source, until self-generation starts and particles get trapped.  

These simple qualitative expectatios have been recently verified using hybrid/Particle-in-cell simulations \cite{Schroer2021,Schroer2022}. The authors also show that, at least for relatively young SNRs, the current density in the form of particles escaping the source can be large enough to excite the non-resonant branch of the streaming instability \cite{Bell2004,AmatoBlasi2009}. 

These modes are excited when the current in the form of particles with momentum $p$ at a given location satisfies the condition \cite{Bell2004}:
\begin{equation}
\Phi_{CR}(>E)E \frac{1}{c} > \frac{B_0^2}{4\pi},   
\end{equation}
where $\Phi(>E)$ is the local flux of particles with energy larger than $E$ at the given location. When particles move ballistically, this condition translates to requiring the energy density of CR being larger than the magnetic energy density in the form of pre-existing magnetic field. In a fully non-linear calculation the flux changes as a consequence of particle transport in the self-generated fields, and eventually the system evolves toward a situation in which equipartition is restored, typically in a larger region, namely a bubble excavated by CRs around the source. The simulations of \cite{Schroer2021,Schroer2022} showed that this chain of events gets starts as expected. However, it is currently not possible to use such hybrid approach to follow the large scale evolution of the system. The bubbles that result from the scenario illustrated above are typically characterized by a low gas density (the gas is evacuated), and a strongly turbulent magnetic structure that manifests itself in the form of a suppression of the diffusion coefficient. 

This line of thought suggests that around any sufficiently powerful source of CRs there should be regions of lower density and suppressed diffusion, making the picture of CR transport in the Galaxy less trivial than often assumed. The effects of these circum-source regions in terms of grammage accumulated by CRs is still matter of investigation, while the residence time of CRs in the bubbles is expected to remain negligible compared with the Galactic confinement time, as inferred from the abundance of $^{10}$Be in CR. 

\subsection{Escape of CRs from the Galaxy and extragalactic sources}
\label{sec:escapeGalaxy}

As we discussed in Sec. \ref{sec:CRprop}, the escape flux of CR protons from the Galaxy is very well known once the assumption of stationarity is employed, which in turn is implemented by using a free escape boundary condition for the transport equation. Using Eq. \ref{eq:escapeFlux}, the flux of escaping particles with energy $>E$ can be written as:
\begin{equation}
    \Phi_{esc}(>E)=\frac{N(>E){\cal R}}{2\Sigma}=\frac{L_{CR}}{2\pi R_{d}^{2}\Lambda}E^{-2},
    \label{eq:escapeFluxCurrent}
\end{equation}
where ${\cal R}$ is the rate of supernovae (or rate of occurrence of other types of sources) and $\Sigma=\pi R_d^2$ is the surface of the Galactic disc of radius $R_d$. In Eq. \ref{eq:escapeFluxCurrent} the integral spectrum contributed by each source is $N(>p)$. If the energy spectrum of the sources is close to $E^{-2}$, then the escape flux can be expressed in terms of the total CR luminosity of the Galaxy, $L_{CR}$ (last equality in Eq. \ref{eq:escapeFluxCurrent}). 

As discussed above, if the current is sufficiently strong that it carries an energy larger than the pre-existing magnetic energy, a non-resonant streaming instability is excited. In the case of the Galaxy, this happens if the following condition on the extragalactic magnetic field is fulfilled \cite{BlasiAmato2019}: 
\begin{equation}
B_{0}\leq B_{sat}\approx 2.2 \times 10^{-8}  L_{41}^{1/2} R_{10}^{-1}\ {\rm G},
\label{eq:Bsat}
\end{equation}
where $L_{41}$ is the CR luminosity of the Galaxy in units of $10^{41}$ erg s$^{-1}$ and $R_{10}$ is the radius of the galactic disk in units of 10 kpc. We assumed $E_{max}=1 PeV$ in computing $\Lambda=\ln(E_{max}/m_p)$.

The excitation of the non-resonant instability leads to a growth of the magnetic field on small scales and eventually to a saturation at $B_{sat}$, as given in the RHS of Eq. \ref{eq:Bsat}. 

The implications of this simple picture are numerous \cite{BlasiAmato2019}: the growth of the field stops when the fluctuations acquire a typical spatial scale comparable with the Larmor radius of the particles in the modified field. While during the exponential growth the instability works on very small scales (much smaller than the gryration radius of the particles dominating the current) and the current is not appreciably modified, its non-linear evolution corresponds to a strong interaction of the field with the current. In other words, the particles are expected to get scattered efficiently, which translates to a small diffusion coefficient. In fact, when this happens the background plasma is set in motion with a speed comparable with the Alfv\'en speed calculated using the amplified magnetic field $B_{sat}$. As discussed by \cite{BlasiAmato2019}, the scenario discussed above suggests that around any galaxy there should be an extended region populated by CRs escaped from the host galaxy but whose propagation becomes advection dominated. Moreover, any galaxy should be surrounded by an extended region with magnetic field of order $B_{sat}$ (with a suitable choice of the parameters of the galaxy, for instance the CR luminosity of that galaxy). These magnetized bubbles are expected to be especially large around sources of UHECR, where a much higher $L_{CR}$ is expected \cite{DAngeloPRL}. For such sources, the propagation of the lower energy part of the spectrum is expected to be heavily affected by the presence of these regions, possibly in the form of a low energy flux suppression. The investigation of these non-linear effects around CR sources is clearly only at the very beginning and much remains to be learned from future studies of these fascinating phenomena. 

\section{Summary}
\label{sec:concl}

Faced with the challenge of writing Lecture Notes on a complex and vast topic such as Cosmic Ray Transport, one has to make a choice on the subjects to cover. Here I chose to focus on the foundations of the problem, and on some of the applications that show how powerful such foundations are. Only in the final part of these Notes I dedicated some space to more modern aspects, so that the reader may get a feeling of how some current research topics are developing. The list of these research topics is, of course, partial and subjective. 

These Notes begin with a short description of the main reasons why we think that CR transport is non trivial, namely that it is not limited to the simple gyration of charged particles in magnetic fields. After this introduction, I proceeded in deriving the basic insights in the problem of describing the motion of a charged particle in a complex magnetic field, starting from the simple unperturbed motion in a regular field. The simple arguments illustrated there introduced the reader to resonant interaction with Alfv\'en waves and the concept of diffusion. I continued with a derivation of the equation describing pitch angle diffusion of a charged particles in a background of Alfv\'en waves, starting from the Vlasov equation. This derivation is a simpler version of a mathematically more cumbersome treatment including modes other than Alfv\'en waves (see for instance \cite{Volk1973,Volk1975}). Although these extensions of the theory are very important for the field, the mathematics can be easily understood once the reader becomes familiar with the simpler version discussed here. 

From the diffusion equation in pitch angle I derived the corresponding transport equation describing spatial diffusion and generalized the result to the case in which the background plasma is in motion with a non uniform velocity profile. This latter case is of the utmost importance for applications of the transport equation to particle acceleration at shocks. 

These results, despite their simplicity, represent very powerful tools to investigate a variety of problems in high energy astrophysics. In fact, very accurate predictions can be obtained using these simple findings or straightforward generalizations (for instance including nuclei and their spallation). More complex treatments are definitely available but often require additional assumptions on some key ingredients. For instance, the diffusion equation discussed here is based on diffusion in one spatial dimension, so that the diffusion coefficient is a scalar. One could write a transport equation accounting for parallel and perpendicular diffusion, but in fact this is rarely done because it requires the knowledge of the connection between turbulence and perpendicular transport, that is not often available: a complete theory of perpendicular transport is not yet available. 

I discuss two applications of the transport equation in space to problems of astrophysical importance: 1) calculation of the spectrum and spatial distribution of CR protons in the Galaxy; 2) particle acceleration at a non relativistic shock. A similar formalism was extended to the case of nuclei and electrons in the Lecture Notes by C. Evoli. In the Lecture Notes by D. Caprioli the problem of diffusive shock acceleration was reconsidered  and extended to its non-linear version, using semi-analytical methods and numerical simulations. 

A primer to the problem of perpendicular particle transport was also presented, but I chose to not adventuring in the world of the numerous models trying to describe such phenomenon.

In the last part of these Notes I briefly discussed selected topics in the non-linear theory of CR transport. The purpose of this part is not to provide all the technical tools necessary to carry out research work, but rather to inform the reader on the evolution of the theory of CR transport and the most impressive implications of its application to CR transport in the Galaxy, around sources and around galaxies. 

The non-linear behaviour of CR transport is now described both using semi-analytical tools and numerical simulations: it turns out to be of the utmost importance for many of the problems discussed here, from diffusive shock acceleration (in the absence of non-linear effects the maximum energy achievable in most astrophysical accelerators is of little relevance) to Galactic CR transport and escape from both galactic and extragalactic sources. One crucial ingredient of these theories is the excitation of plasma instabilities due to CR streaming. The linear growth of these instabilities has been covered in the Lecture Notes of A. Marcowith. 

Several topics have only been mentioned, although I consider them as priorities and active investigation is ongoing. One such topic is the description of the particle transport in MHD turbulence, most notably fast magnetosonic modes: at energies above few hundred GeV this process has been proposed to be the chief mechanism of particle scattering \cite{YL2002,YL2004,YL2008}, but many concerns exist on whether the formation of weak shocks may affect this process and whether this scattering can in fact reproduce the observed shapes of CR spectra \cite{Kempski2022}. The problem of transport in MHD turbulence is tightly connected with the problem of cascading of such turbulence from large scales and random walk of magnetic field lines, which in turn provides a strong contribution to perpendicular CR transport.  

Another topic that would deserve a dedicated section is the discussion of anisotropic diffusion: there are currently many versions of the NLGCT that attempt to describe particle transport perpendicular to the ordered magnetic field lines (for given parallel transport), but the comparison of these models' predictions with the results of test particle simulations seems successful only for some selected simple models of turbulence. 

Finally, I want to mention that the non-linear interaction of CRs with the environment in which they propagate, known as CR feedback, is central to numerous problems in Galactic and extragalactic astrophysics, from the launching of CR driven winds and regulation of the star formation rate to the excavation of bubbles around sources of CRs. The problem of CR feedback is discussed in the Lecture Notes by E. Zweibel.   

%\bibliographystyle{mnras}
%\bibliography{biblio}

\end{document}